%% file: aspmv.tex
\RequirePackage{fix-cm}
\documentclass[twocolumn]{svjour3}          

\input{preamble}

\begin{document}

\title{Characterizing Scalability of Sparse Matrix-Vector Multiplications on Phytium FT-2000+ Many-cores}

\titlerunning{Characterizing Scalability of SpMV on Phytium FT-2000+ Many-cores}        


\author{Donglin Chen \and
Jianbin Fang\and
Chuanfu Xu\and
Shizhao Chen\and
Zheng Wang}

\authorrunning{D. Chen et al.} 

\institute{D. Chen, J. Fang, C. Xu, S. Chen are with
              College of Computer Science, National University of Defense Technology, Changsha 410073, China. \\
              \email{\{chendonglin14, j.fang, xuchuanfu, chenshizhao12\}@nudt.edu.cn}
           \and
           \\ Z. Wang is with
              University of Leeds, United Kingdom. \\
			\email{z.wang5@leeds.ac.uk}	
}

\date{}

\maketitle

\begin{abstract}
Understanding the scalability of parallel programs is crucial for software optimization and hardware architecture design. As HPC hardware
is moving towards many-core design, it becomes increasingly difficult for a parallel program to make effective use of all available
processor cores. This makes scalability analysis increasingly important. This paper presents a quantitative study for characterizing the
scalability of sparse matrix-vector multiplications (SpMV) on Phytium FT-2000+, an ARM-based HPC many-core architecture. We choose SpMV
as it is a common operation in scientific and HPC applications. Due to the newness of ARM-based many-core architectures, there is little
work on understanding the SpMV scalability on such hardware design. To close the gap, we carry out a large-scale empirical evaluation
involved over 1,000 representative SpMV datasets. We show that, while many computation-intensive SpMV applications contain extensive
parallelism, achieving a linear speedup is non-trivial on Phytium FT-2000+. To better understand what software and hardware parameters
are most important for determining the scalability of a given SpMV kernel, we develop a performance analytical model based on the
regression tree. We show that our model is highly effective in characterizing SpMV scalability, offering useful insights to help
application developers for better optimizing SpMV on an emerging HPC architecture.


\end{abstract}

\input{spmv_intro}

\input{spmv_background}
\input{spmv_setup}

\input{spmv_results}

\input{spmv_analysis}
\input{spmv_advise}
\input{spmv_relatedwork}

\input{spmv_conclusion}

\begin{acknowledgements}
This work was partially funded by the National Key R\&D Program of China under Grant agreements 2017YFB0202003,
 the National Natural Science Foundation of China under grant agreements 61602501,
61772542, and 61872294; and the Royal Society International Collaboration Grant (IE161012). For any correspondence,
please contact Jianbin Fang (Email: j.fang@nudt.edu.cn) and Chuanfu Xu (Email: xuchuanfu@nudt.edu.cn).
\end{acknowledgements}

 \bibliographystyle{spbasic}
\bibliography{aspmv,zheng}

\end{document}

%% file: preamble.tex
\usepackage{makeidx}  
\usepackage{booktabs} 
\usepackage{amsfonts}
\usepackage[dvips]{graphicx}
\usepackage[tight,footnotesize]{subfigure}
\usepackage{listings}
\usepackage{cite}
\usepackage{url}
\usepackage{pifont}
\usepackage{flushend}
\usepackage{algorithm}
\usepackage{algpseudocode}
\usepackage[dvipsnames]{xcolor}
\usepackage{float}
\usepackage{amsmath,xparse}
\usepackage{color, colortbl}

\smartqed  
\usepackage{graphicx}

\usepackage{multirow}
\usepackage{booktabs}
\usepackage{ulem}
\usepackage{color, soul}
\soulregister\cite7 
\soulregister\citep7 
\soulregister\citet7 
\soulregister\ref7 
\soulregister\pageref7 

\usepackage{setspace}
\usepackage[square,sort,comma,numbers]{natbib}
\usepackage{booktabs}
\usepackage{footnote}

\definecolor{Gray}{gray}{0.9}
\newcommand{\cparagraph}[1]{\vspace{1.5mm}\noindent\textbf{#1.}}

\makeatletter
\def\therule{\makebox[\algorithmicindent][l]{\hspace*{.5em}\vrule height .75\baselineskip depth .25\baselineskip}}%

\newtoks\therules
\therules={}
\def\appendto#1#2{\expandafter#1\expandafter{\the#1#2}}
\def\gobblefirst#1{
  #1\expandafter\expandafter\expandafter{\expandafter\@gobble\the#1}}%
\def\LState{\State\unskip\the\therules}
\def\pushindent{\appendto\therules\therule}%
\def\popindent{\gobblefirst\therules}%
\def\printindent{\unskip\the\therules}%
\def\printandpush{\printindent\pushindent}%
\def\popandprint{\popindent\printindent}%

\algdef{SE}[WHILE]{While}{EndWhile}[1]
  {\printandpush\algorithmicwhile\ #1\ \algorithmicdo}
  {\popandprint\algorithmicend\ \algorithmicwhile}%
\algdef{SE}[FOR]{For}{EndFor}[1]
  {\printandpush\algorithmicfor\ #1\ \algorithmicdo}
  {\popandprint\algorithmicend\ \algorithmicfor}%
\algdef{S}[FOR]{ForAll}[1]
  {\printindent\algorithmicforall\ #1\ \algorithmicdo}%
\algdef{SE}[LOOP]{Loop}{EndLoop}
  {\printandpush\algorithmicloop}
  {\popandprint\algorithmicend\ \algorithmicloop}%
\algdef{SE}[REPEAT]{Repeat}{Until}
  {\printandpush\algorithmicrepeat}[1]
  {\popandprint\algorithmicuntil\ #1}%
\algdef{SE}[IF]{If}{EndIf}[1]
  {\printandpush\algorithmicif\ #1\ \algorithmicthen}
  {\popandprint\algorithmicend\ \algorithmicif}%
\algdef{C}[IF]{IF}{ElsIf}[1]
  {\popandprint\pushindent\algorithmicelse\ \algorithmicif\ #1\ \algorithmicthen}%
\algdef{Ce}[ELSE]{IF}{Else}{EndIf}
  {\popandprint\pushindent\algorithmicelse}%
\algdef{SE}[PROCEDURE]{Procedure}{EndProcedure}[2]
   {\printandpush\algorithmicprocedure\ \textproc{#1}\ifthenelse{\equal{#2}{}}{}{(#2)}}%
   {\popandprint\algorithmicend\ \algorithmicprocedure}%
\algdef{SE}[FUNCTION]{Function}{EndFunction}[2]
   {\printandpush\algorithmicfunction\ \textproc{#1}\ifthenelse{\equal{#2}{}}{}{(#2)}}%
   {\popandprint\algorithmicend\ \algorithmicfunction}%
\makeatother

\makesavenoteenv{table}
\makesavenoteenv{tabular}

%% file: spmv_intro.tex
\section{Introduction}
Multi-core and many-core architectures offer the potential of delivering scalable performance through parallelism. Realizing such potential is,
however, not trivial due to multiple factors, including available application parallelism, limited working sets, and communication
overheads. Among these factors, the share memory resources, such as shared caches, is often a performance bottleneck for many application domains due to memory
contention~\cite{DBLP:conf/ics/LiuLS08}.

The memory bandwidth is increasingly becoming a limiting factor for the high-performance computing (HPC) domain. On the one hand, there are
more and more processor cores that are integrated into a single chip, to provide more computation power. On the other hand, using a
larger number of processor cores is likely to raise memory contention and increase the pressure on the memory bus. As a result, it is not
always beneficial to use a large number of cores even if abundant parallelism is available~\cite{Gupta2012Evaluating}. To unlock the
potential of multi- and many-core architectures and to justify the further specialization of processor design, it is important to
understand the impact of
the shared memory resources on application scalability.

In this paper, we present a quantitative approach to characterize the scalability of sparse matrix-vector multiplications (SpMV) on HPC
many-core architectures. SpMV is one of the most common operations in scientific and HPC applications~\cite{DBLP:conf/ics/SedaghatiMPPS15}. 
It is highly challenging to optimize SpMV on parallel
architectures~\cite{DBLP:journals/pc/WilliamsOVSYD09}, due to several reasons like irregular indirect data accessing, sensitivity to the
sparsity pattern of the input matrix, and the subtle interaction of the matrix storage format, the problem size, and hardware. While there
is considerable work on finding the right sparse matrix storage
format~\cite{DBLP:conf/sc/BellG09,osti_7093021,DBLP:conf/ics/0002V15,DBLP:conf/hipeac/MonakovLA10,DBLP:journals/siamsc/KreutzerHWFB14},
little effort has been spent on characterizing and understanding the scalability of SpMV on multicore architectures. As the HPC hardware is
firmly moving towards many-core design, it is crucial to know when it is beneficial to use the available cores and how the SpMV performance
will scale as we increase the number of cores to use.

Our work specifically targets the ARMv8-based Phytium \texttt{FT-2000+} many-core architecture. Because ARM-based processors are emerging
as an interesting alternative building block for HPC systems~\cite{DBLP:conf/hotchips/Zhang15, DBLP:conf/hotchips/Stephens16, DBLP:conf/ispass/LaurenzanoTCJWC16},
it is important to understand how the hardware microarchitecture design affects the SpMV scalability. Having such knowledge is useful not
only for better utilizing the computation resources, but also for justifying a further increase in the processor core provision on a single
chip.

In this work, we conduct a comprehensive evaluation and analysis to study the scalability of SpMV on the latest \texttt{FT-2000+} many-core.
Our study mainly targets the Compressed Sparse Row (CSR) storage format. We choose CSR because it is a widely used representative storage
format for sparse matrices in scientific computing. Since there are many variations of the CSR format, our optimization has great practical
significance and can easily be extended to other CSR-extended formats.

Our experiment shows that despite many SpMV applications contain extensive parallelism, they often fail to achieve a linear speedup on
\texttt{FT-2000+}. To character what affects the scalability of SpMV, we collect extensive profiling information (through hardware
performance counters) from a large-scale experiment involved over 1,000 representative sparse datasets. With this extensive set of
profiling data in place, we develop a regression-tree based analytical model to capture what information is useful for reasoning about the
scalability of SpMV. We show that our analytical model is highly accurate in revealing what affects the SpMV scalability on
\texttt{FT-2000+}. We demonstrate that our model can provide useful insights to guide the application developers to better optimize SpMV on
an emerging ARMv8-based many-core architecture.

To summarize, this paper makes the following contributions. It is the first to

\begin{itemize}
	\item characterize the scalability performance of SpMV on \texttt{FT-2000+}, an emerging ARMv8-based many-core architecture for HPC;
	
	\item use machine learning techniques to correlate and analyze how hardware micro-architecture features affect the SpMV scalability.

	\item show how machine learning can be used to develop a performance profiling tool to guide the optimization of SpMV on ARM-based HPC architectures.
	
\end{itemize}

%% file: spmv_background.tex
\section{Background and Motivation}
In this section, we first introduce the SpMV and its sparse matrix storage formats and then explain the motivation of this work.

\subsection{Sparse Matrix-Vector Multiplication}
\begin{figure}[!t]
\centering
\includegraphics[width=0.45\textwidth]{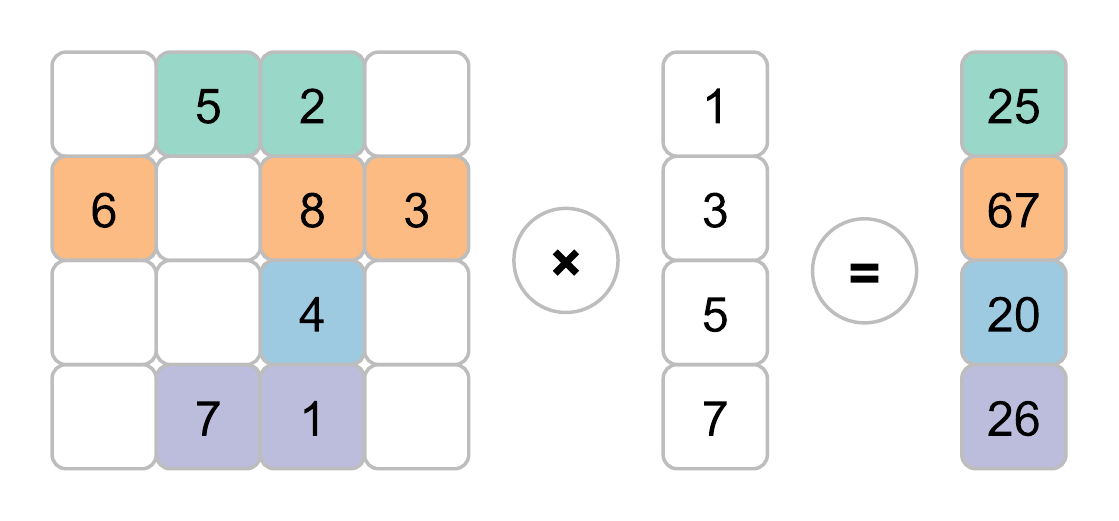}
\caption{A simple example of SpMV with a $4 \times 4$ matrix ($nnz=8$) by a $4 \times 1$ vector. The product of this SpMV is a $4 \times 1$ vector.}
\label{fig:spmv_example}
\end{figure}

A SpMV operation can be defined as $\mathbf{y}=\mathbf{Ax}$
where the input is a sparse matrix $\mathbf{A}$ ($m \times n$) and
a dense vector $\mathbf{x}$ ($n \times 1$), and the output is a dense vector $\mathbf{y}$ ($m  \times 1$).
Figure~\ref{fig:spmv_example} shows an illustrative example of SpMV, where
$\mathbf{m}=\mathbf{n} = 4$, and the nonzeros $nnz=8$.

\subsection{Sparse Matrix Storage Formats\label{sec:format}}
In our work, we mainly consider the SpMV based on CSR, the most commonly used format for storing sparse matrices, and its improved counterpart, CSR5~\cite{DBLP:conf/ics/0002V15}. The example matrix mentioned above in these two formats is shown in Table~\ref{tbl:matrix_format}.
%
%

	

\begin{table*}[!t]
\caption{The sparse matrix storage formats targeted in this work and the corresponding data structures for the example shown in Figure~\ref{fig:spmv_example}.}
\label{tbl:matrix_format}
\scriptsize
\begin{center}
\scalebox{1.20}{
\begin{tabular}{crcr} \toprule
\textbf{Representation} & \textbf{Specific Values}  
\\
\midrule
\rowcolor{Gray}
 CSR  & \begin{tabular}{@{}r@{}}$ptr=[0, 2, 5, 6, 8]$ \\ $indices=[1, 2, 0, 2, 3, 2, 1, 2]$ \\ $data=[5, 2, 6, 8, 3, 4, 7, 1]$\end{tabular}

\\

CSR5  & \begin{tabular}{@{}r@{}}$ptr=[0, 2, 5, 6, 8]$  $tile\_ptr=[0, 1, 4]$\\
$tile\_des: bit\_flag=[T, T, F, F | T, T, T, F],$\\ $y\_off=[0, 1 | 0, 2], seg\_off=[0, 0 | 0, 0]$ \\
$indices=[1, 0, 2, 2 | 3, 1, 2, 2]$ \\  $data=[5, 6, 2, 8 | 3, 7, 4, 1]$\end{tabular}

\\
\rowcolor{Gray}
\bottomrule
\end{tabular}}
\end{center}

\end{table*}

\cparagraph{CSR} The \textit{compressed sparse row} (CSR) format explicitly stores column indices and nonzeros in arrays \texttt{indices}
and \texttt{data}, respectively. It uses a vector \texttt{ptr}, which points to row starts in \texttt{indices} and \texttt{data}, to query
matrix values. The length of \texttt{ptr} is $n\_row + 1$, where the last item is the total number of the nonzero elements of the matrix.

\cparagraph{CSR5} The CSR5 format aims to obtain a good load balance for matrix value queries~\cite{DBLP:conf/ics/0002V15}. It achieves this by partitioning all nonzero elements into multiple 2-dimensional tiles of the same size. 
corresponding to the width and the height of the title respectively. Later in this paper, we show how CSR5 gain better scalability than CSR by more uniform and reasonable task assignment in multi-threaded SpMV.

\subsection{Motivation}
\begin{figure}[!t]
\centering
\includegraphics[width=0.45\textwidth]{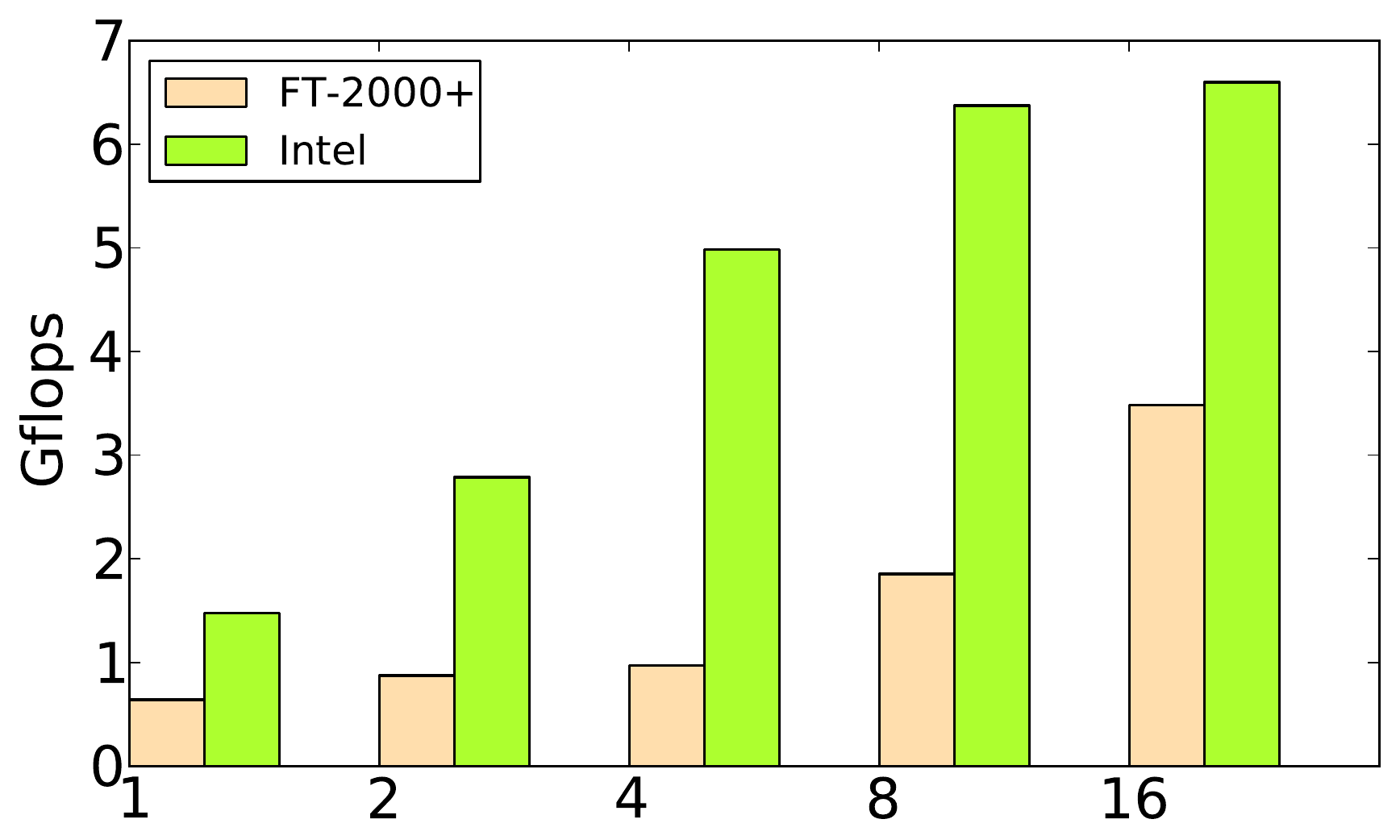}
\caption{Performance comparison of SpMV on two multicore processors.
The x-axis represents the number of threads and the y-axis represents the obtained performance (in Gflops).}
\label{fig:motivation}
\end{figure}

We run the multi-threaded SpMV in CSR on a \texttt{x86}-based Xeon multi-core (\texttt{Intel Xeon E5-2692}) and a \texttt{ARMv8}-based Phytium multi-core (\texttt{FT-2000+}).
Figure~\ref{fig:motivation} shows the SpMV performance for the \texttt{bone010} dataset when using threads ranging from 1 to 16.

We observe that, on \texttt{Xeon} the speedup increases linearly when using 1 thread upto 4 threads,
while the performance increase is very slight when using furthermore threads.
At this moment, the SpMV performance on \texttt{Xeon} is limited by the off-chip memory accesses.
By contrast, the SpMV scalability is rather different on \texttt{FT-2000+}.
We see a very slight performance increase when using 1, 2, and 4 threads.
Thereafter, we notice a quasi-linear speedup until using 16 threads.
We believe that these performance behaviours are determined by the interactions of the SpMV code, the input sparse matrix, and the underlying micro-architecture.
In this work, we will look into the factors which impact the SpMV scalability on \texttt{FT-2000+}.

Given that the performance `odds' appear when using fewer than 8 threads, we will focus on scalability characterization on a core-group within a panel of \texttt{FT-2000+} (see Figure~\ref{fig:ft2000_architecture} and Section~\ref{sec:setup}).

%% file: spmv_setup.tex

\section{Experimental Setup} \label{sec:setup}

In this section, we will introduce the hardware platforms, the installed system software, and the datasets used in this work.

\begin{figure}[!t]
\centering
\includegraphics[width=0.5\textwidth]{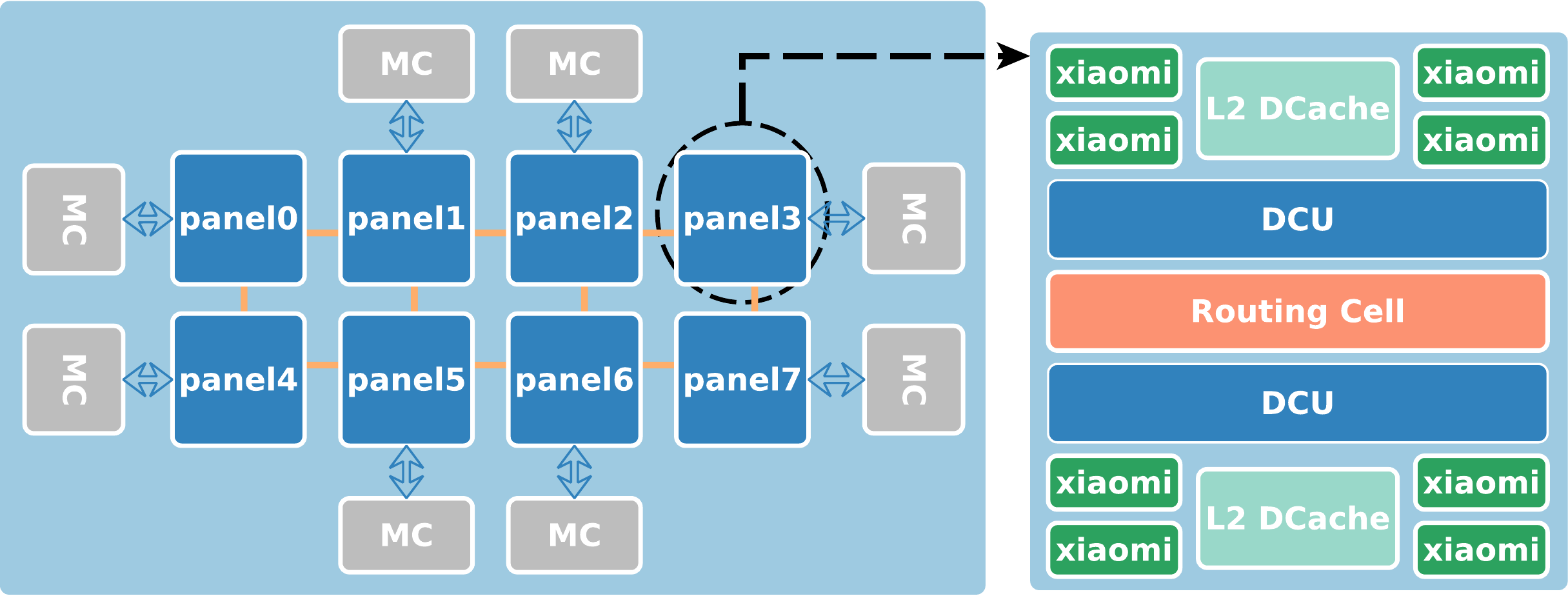}

\caption{A high-level view of the \texttt{FT-2000+} architecture. Processor cores
	are groups into panels (left) where each panel contains eight ARMv8 based
	Xiaomi cores (right).} \label{fig:ft2000_architecture}

\end{figure}

\cparagraph{Hardware Platforms} As depicted in Figure~\ref{fig:ft2000_architecture}, \texttt{FT-2000+} integrates 64 ARMv8 based Xiaomi cores. Its \texttt{Mars II} microarchitecture
offers a peak performance of 588.8 Gflops for double-precision operations, with a maximum power consumption of 96 Watts. The CPU chip has eight panels with eight 2.3GHz cores per panel. Each core has a private 32KB L1 data cache, and a 2MB L2 cache is
shared among four cores (\texttt{core-group}). The panels are connected through two directory control units (\texttt{DCU})~\cite{ft2000plus}.

\cparagraph{Systems Software} We run a customized Linux OS with Linux Kernel v4.4 on \texttt{FT-2000+}. For compilation, we
use gcc v6.4.0 with the ``-O3" compiler option. We use the OpenMP threading model, using 1-4 threads on \texttt{FT-2000+}.

\cparagraph{Datasets} We use 1008 square matrices (with a total size of 80 GB) from the SuiteSparse matrix
collection~\cite{DBLP:journals/toms/DavisH11}. The number of nonzero elements of the matrices ranges from 100K to 200M~\cite{DBLP:conf/ppopp/LiuHLT18}. The dataset includes
both regular and irregular matrices, covering domains from scientific computing to social networks~\cite{DBLP:journals/pc/0002V15}.

%% file: spmv_results.tex
\section{SpMV Scalability Results and Modelling}

In this section, we show the overall scalability performance of SpMV.
To identify the impacting factors of SpMV scalability on \texttt{FT-2000+}, we build a regression-tree based model, which automated relates features to speedup (normalized to a single thread).
We use key features collected from hardware performance events and the input sparse matrix datasets.

\begin{figure}[!t]
	\centering
	\includegraphics[width=0.5\textwidth]{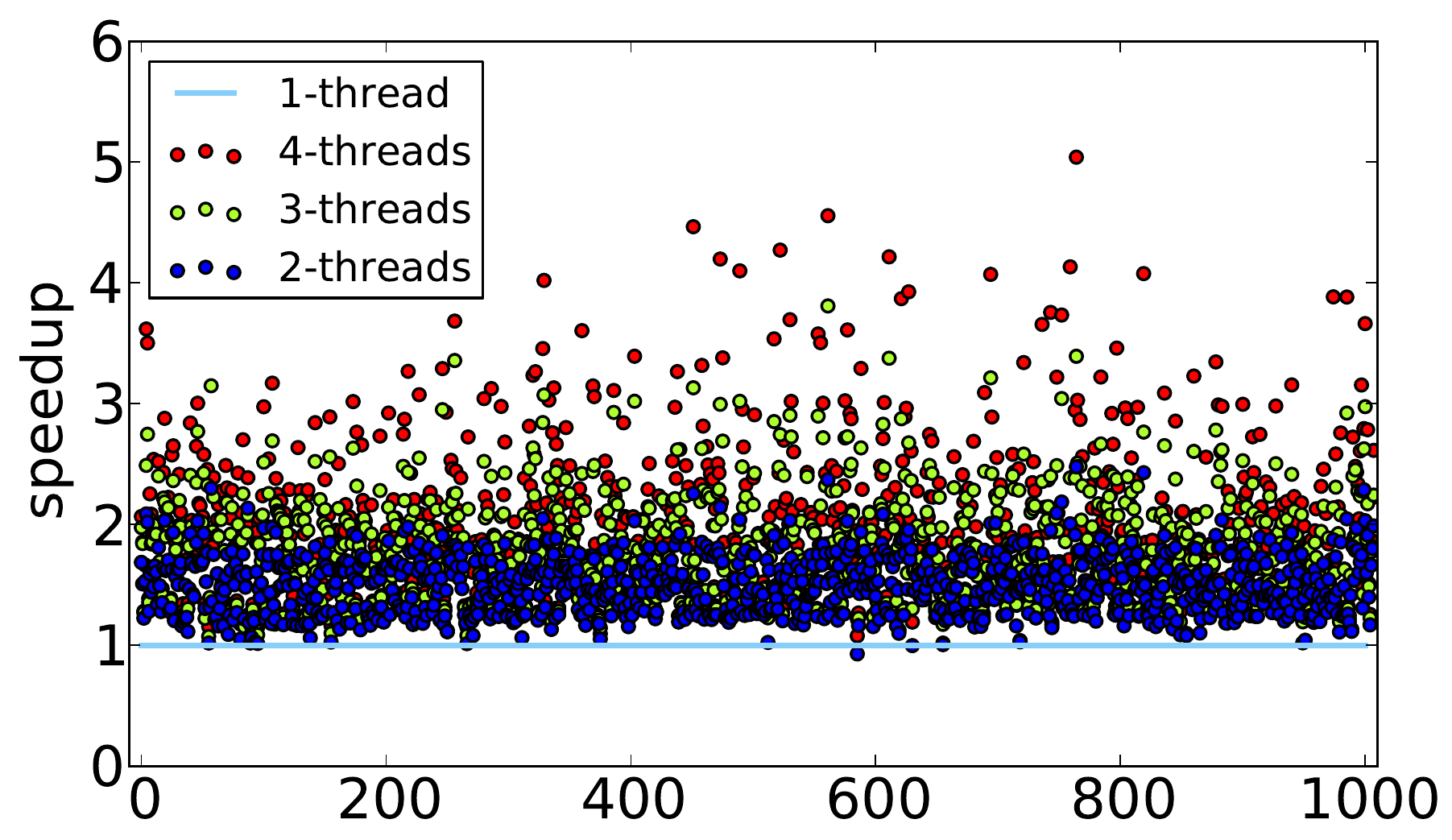}
	\caption{The overall speedup of SpMV in 1-4 threads on \texttt{FT-2000+}.
		The x-axis labels different sparse matrices.}
	\label{fig:overall speedup}

\end{figure}


\subsection{Overall Performance Results}

Figure~\ref{fig:overall speedup} shows the overall speedup of SpMV with 1-4 threads on a core-group of \texttt{FT-2000+}.
The x-axis represents different sparse matrices.
Although the achieved speedup for most matrices increases over the number of threads, we note the performance is far less than the linear speedup.
Most speedup numbers lie between 1 and 2, and a very small portion of numbers are beyond that.
Also, the obtained speedup is hyper-linear for some datasets.
This is because the dataset is so small that it can be hold within the shared L2 data cache.
Table~\ref{tbl:average_speedup} shows a statistical profile of the average speedup when using multiple threads (normalized to that of a single thread).
We see that the average performance of SpMV only improves 50\% from 1 thread to 2 threads and does even not double the number when using 4 threads.
The scalability of SpMV on \texttt{FT-2000+} is far less than our expectation, which motivates us
to identify the impacting factors behind it.

\begin{table}[!t]
	\centering
	\caption{The average speedup(x) of SpMV with multi-threads over a single-thread.}
	\label{tbl:average_speedup}
	\scalebox{0.80}{
		\begin{tabular}{@{}llllll@{}}
			\toprule
			 \rowcolor{white}\textbf{\#threads} &\textbf{1}         & \textbf{2}        & \textbf{3}       & \textbf{4}                 \\
			\midrule
			\textbf{speedup}  & 1.0x 		  & 1.50x			& 1.77x		& 1.93x			\\ \bottomrule
		\end{tabular}}
		
	\end{table}
	
\subsection{Scalability Modelling}

To find the impacting factors for scalability, we use an empirical approach to manually analyze the performance behaviours.
As an alternative, we use a machine-learning based approach to build a model and then let the model tell us which feature plays a role in scaling SpMV on \texttt{FT-2000+}.
Instead of hand-crafting an analytical model that requires expert insight into low-level hardware details, we employ machine learning techniques to
automatically learn the correlation between features and the SpMV (speedup) performance.

Building and using the regression tree model follows three main steps:
(i) generate training data, (ii) train a regression model, and (iii) find the factors with a large weight.
Given that our model is used as a tool for analysis rather than for predicting the speedup of SpMV, we make the best use of the collected data by
selecting 90\% samples for training, instead of the usual (80\%, 20\%) data splitting between model training and model testing.

\subsubsection{Collecting Training Data}
To generate training data for our model, we used 1008 sparse matrices from the SuiteSparse
matrix collection.
We run the CSR-based SpMV
a number of times
until the gap of the upper and lower confidence bounds is smaller than 5\%
under a 95\% confidence interval setting.
The code is run with 1, 2, 3, and 4 threads, with each pinned to a fixed core.
We then record the SpMV execution time for computing speedup (normalized to a single thread) and
obtain hardware performance counters by using \texttt{PAPI} (Performance Application Programming Interface~\cite{DBLP:conf/ptw/TerpstraJYD09}) for each training sample.
As the last step, we collect key values for each input dataset to capture its features.

Table~\ref{tbl:features} shows our selected features from both sparse matrix structure and hardware events.
These important matrix features introduced in~\cite{DBLP:conf/icpp/BenatiaJWS16} are proved to be effective in capturing the spatial patterns of the matrix.
The raw hardware counters we collected are related to performance~\cite{DBLP:conf/sc/MagniDO13}.
To improve the model performance, we calculate a set of derived features based on raw counter values and use them as the input of the model.
There are two customized features:\texttt{L2\_DCMR\_change} and \texttt{job\_var}.
The former indicates the changes of \texttt{L2\_DCMR} from one to four threads.
As for the \texttt{L2\_DCMR} with four threads,
we use the \texttt{L2\_DCMR} on the slowest thread instead of the total one;
the \texttt{job\_var} represents the degree of nonzero distribution imbalance across threads (the theoretical value is 0.25 for 4 threads).

\begin{table*}[!t]
	\centering
	\caption{The selected features and their descriptions.}
	\label{tbl:features}
	\scriptsize
	\begin{tabular}{lll}
	
		\toprule


		&\textbf{Features}   & \textbf{Description}   \\ \midrule
		\rowcolor{Gray}& n\_rows   & number of rows    \\
		\rowcolor{Gray}& nnz\_max  & maximum \# nonzeros per row        \\
		\rowcolor{Gray}& nnz\_avg  & average \#nonzeros per row    \\
		\rowcolor{Gray}\multirow{-4}{*}{matrix features}& nnz\_var  & variance \# nonzeros per row   \\
		& L2\_DCM  & L2 data cache misses  \\
		&L2\_DCA  &	L2 data cache accesses \\
		&L1\_DCM  & L1 data cache misses   \\
		&L1\_DCA  &	L1 data cache accesses \\
		&FR\_INS  & floating point instructions executed   \\
		&TOT\_INS  & total instructions executed \\
		\multirow{-7}{*}{\shortstack{raw \\ hardware counters}}&TOT\_CYC  &  total cycles   \\
		\rowcolor{Gray}& L1\_DCMR  &	L1 data cache miss rate \\
		\rowcolor{Gray}& L2\_DCMR  &	L2 data cache miss rate \\
		\rowcolor{Gray}& IPC  &	instructions per cycle\\ 	
		\rowcolor{Gray}& L2\_DCMR\_change  &the change of L2\_DCMR \\
		\rowcolor{Gray}\multirow{-5}{*}{\shortstack{derived \\ hardware counters}}& job\_var  & maximum \# allocated nnz ratio per thread   \\
		\bottomrule
	\end{tabular}
	
\end{table*}
\subsubsection{Building The Model}
For simplicity, we only use performance counters collected when using one thread and four threads.
The achieved speedup and the corresponding feature set is taken as the input
of the supervised learning algorithm built in \texttt{scikit-learn}.
The learning algorithm tries to find a correlation between the features, performance values and achieved speedups.
The output of this training process
is a regression-tree based model, which helps to reveal the factors that affect SpMV scalability.

\subsubsection{Identifying the Impacting Factors \label{sec:factors}}
\begin{figure}[!t]
	\centering
	\includegraphics[width=0.50\textwidth]{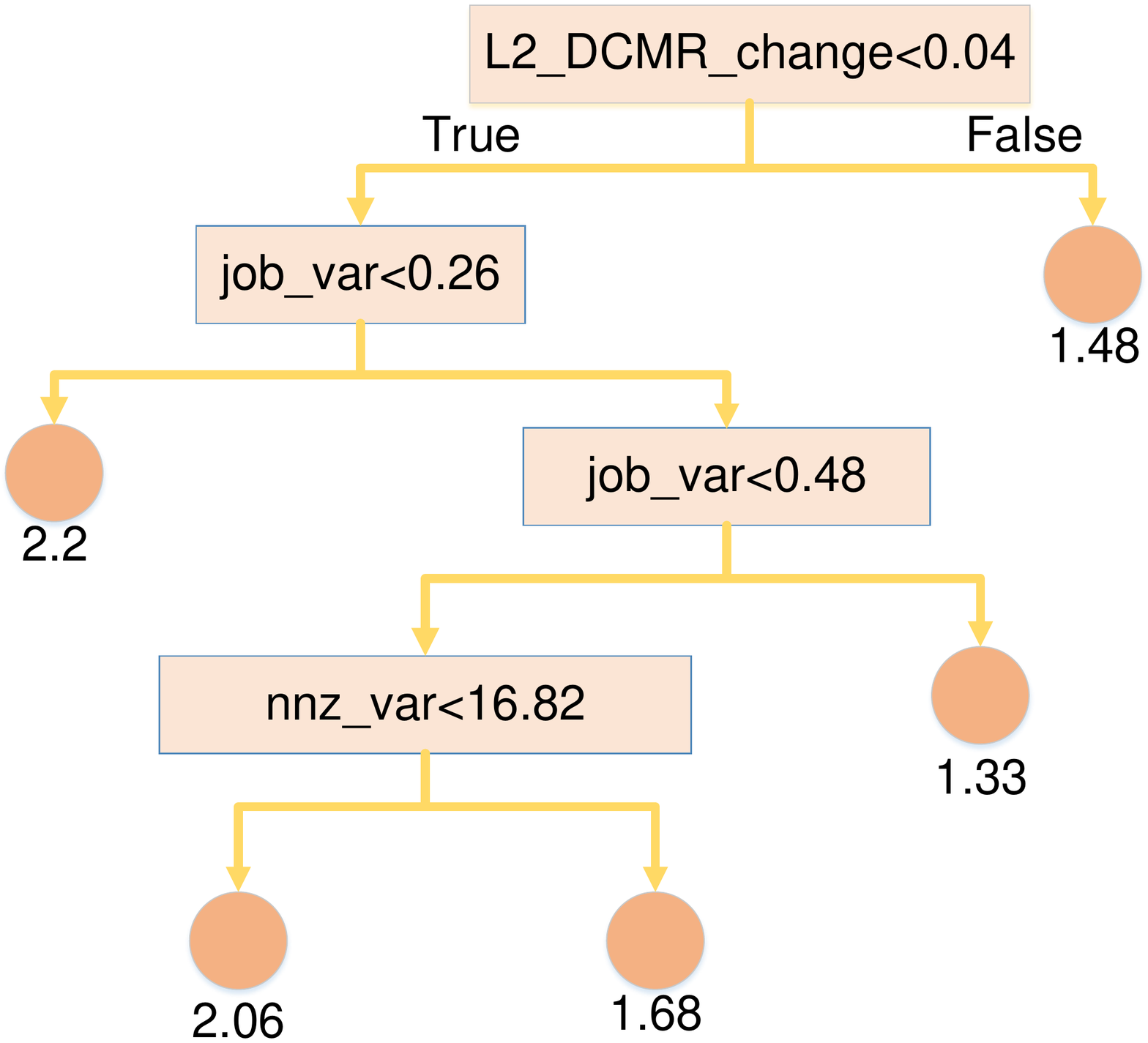}

	\caption{A tree picked from the regression forests intuitively
		shows how the factors impact the speedup of SpMV. }
	\label{fig:tree}

\end{figure}

By using
the \texttt{feature importance} module of \texttt{scikit-learn} for the new-built regression tree model~\cite{scikitlearn},
we can obtain the top three factors that most affect
the SpMV scalability: \textit{the nonzero allocation}, \textit{the shared L2 cache}, and \textit{the nnz variance across rows}, where \textit{nnz} denotes the number of nonzero.
Figure~\ref{fig:tree} shows how these factors
impact the SpMV speedup.
In the next section, we will give a detailed analysis of the scalability with our trained model.

%% file: spmv_analysis.tex
\section{Scalability Analysis, Insights and Optimizations}


In this section,
we first examine how individual factor suggested by the model (Section~\ref{sec:factors}) impacts the SpMV speedup.
We then conduct an in-depth
analysis of how the factors have an impact on the SpMV scalability with four representative matrices.
We choose the four datasets because their speedups
are mainly limited by separate factors.
At last, we introduce several potential optimizations inspired by the scalability results.

\begin{figure*}[!t]
	\centering
	\subfigure[job\_var]{\label{fig:job-var}\includegraphics[width=0.45\textwidth]
		{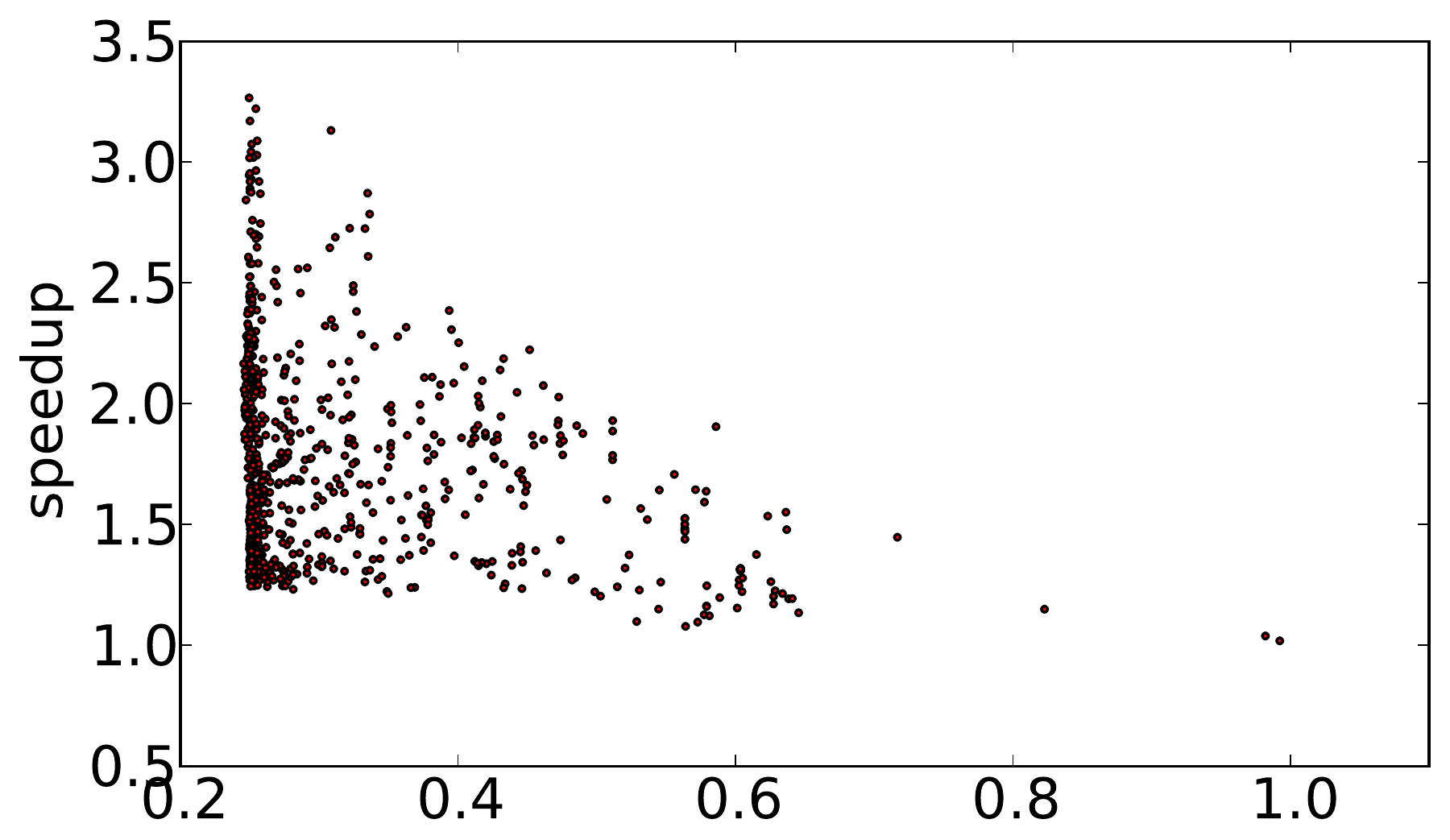}}
	\subfigure[job\_var]{\label{fig:job-var-bar}\includegraphics[width=0.45\textwidth]
		{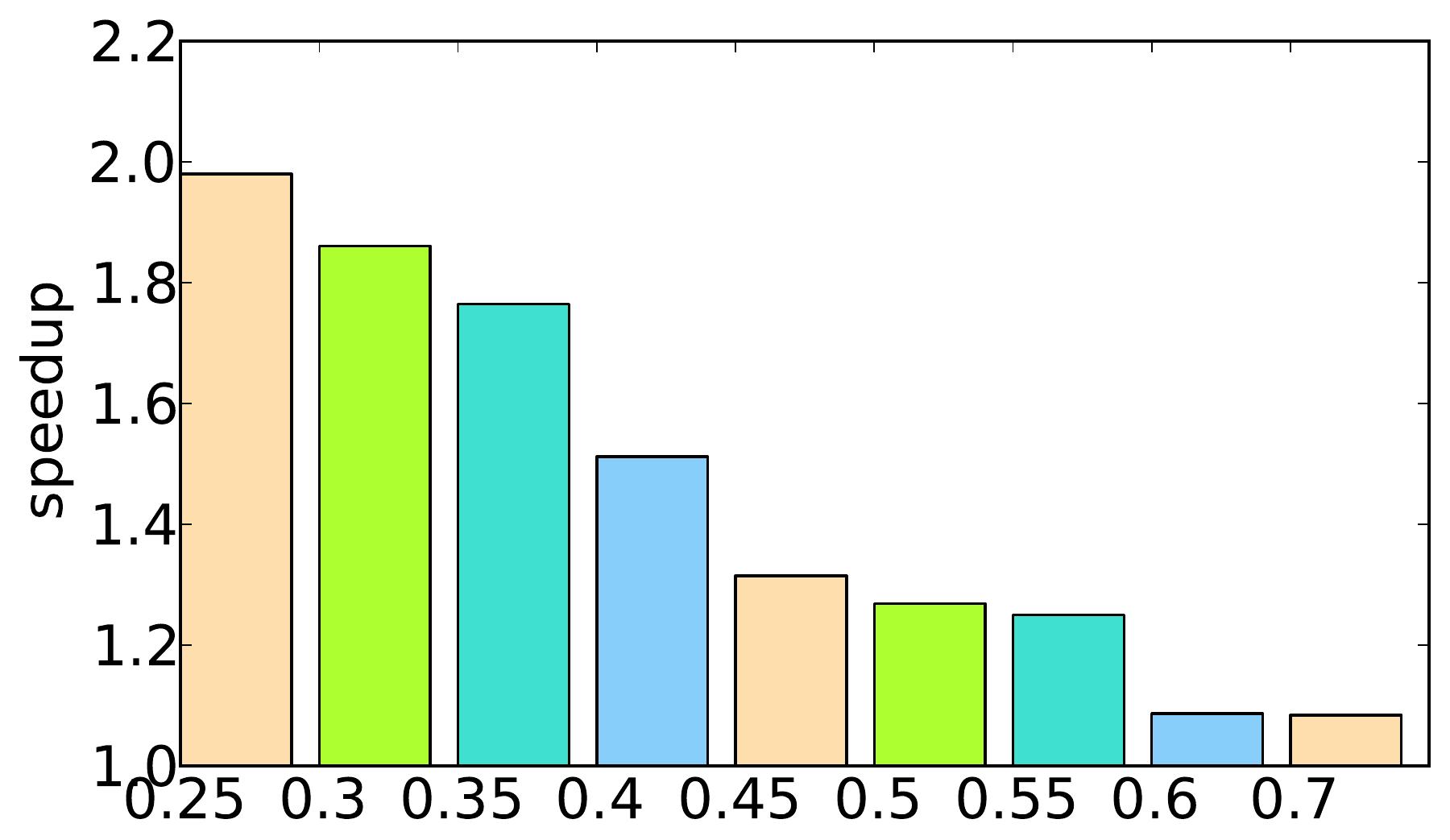}}
	\subfigure[L2\_DCMR change]{\label{fig:L2-cache}\includegraphics[width=0.45\textwidth]
		{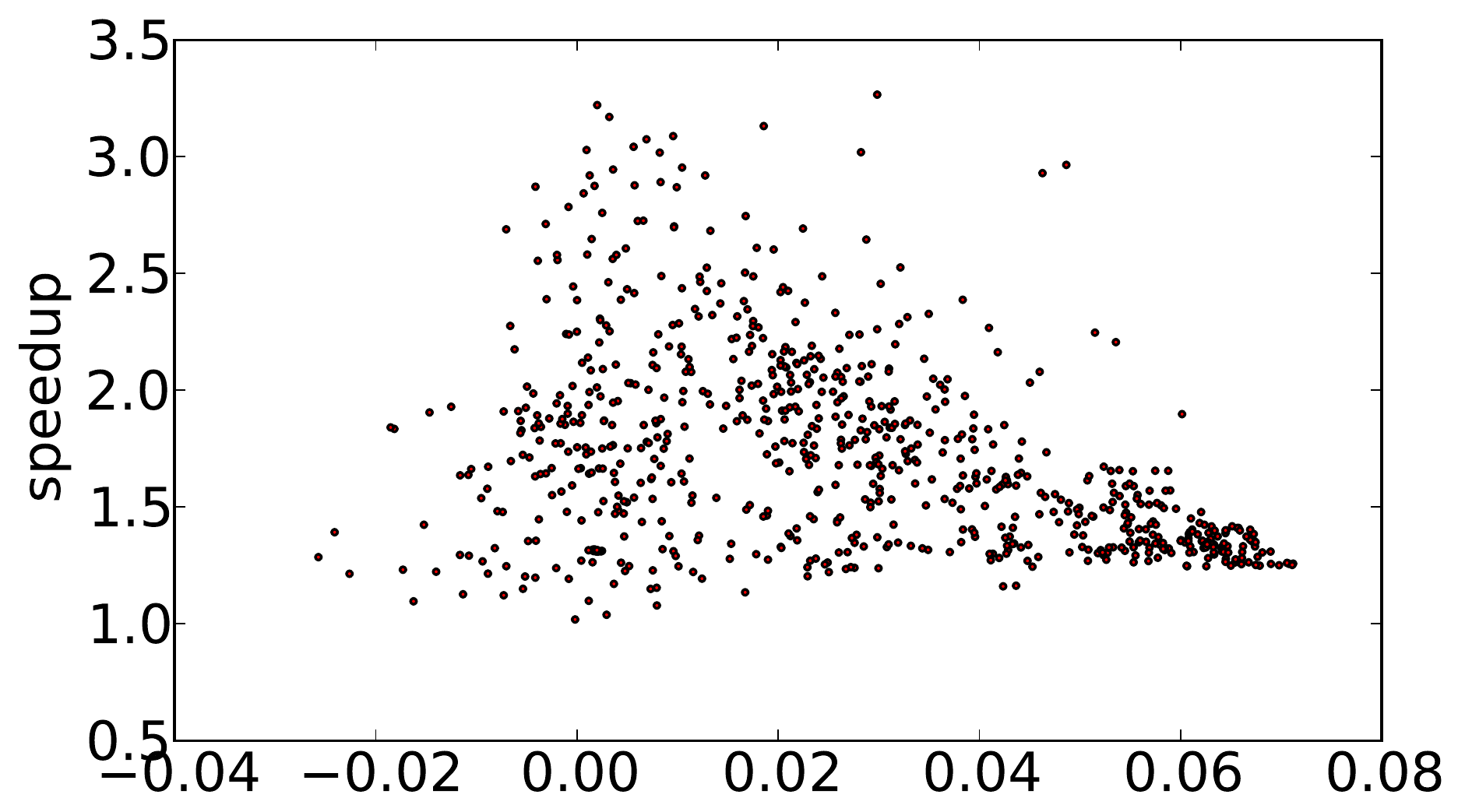}}
	\subfigure[L2\_DCMR change]{\label{fig:L2-cache-bar}\includegraphics[width=0.45\textwidth]
		{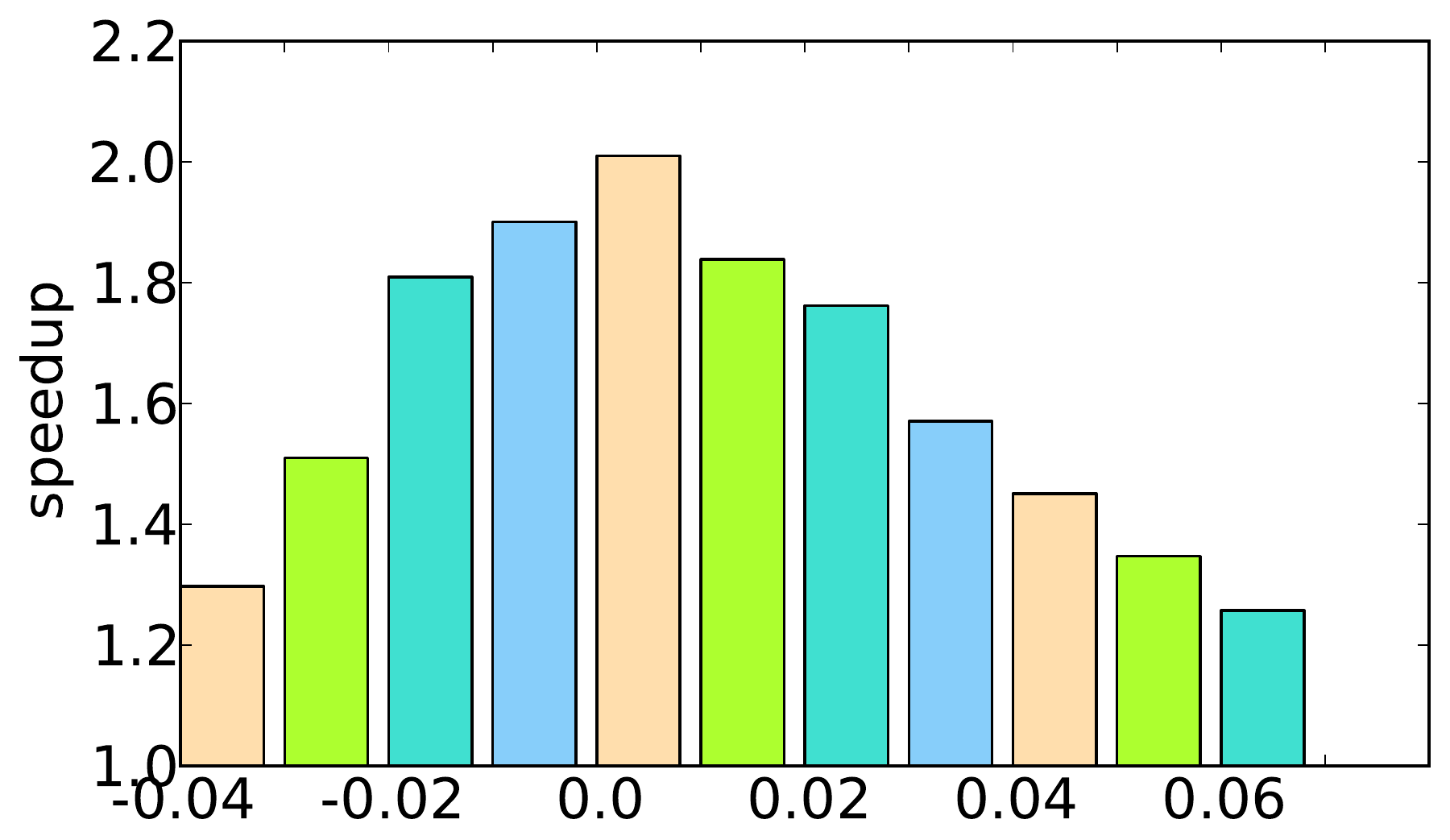}}
	\subfigure[nnz\_var]{\label{fig:nnz-var}\includegraphics[width=0.45\textwidth]
		{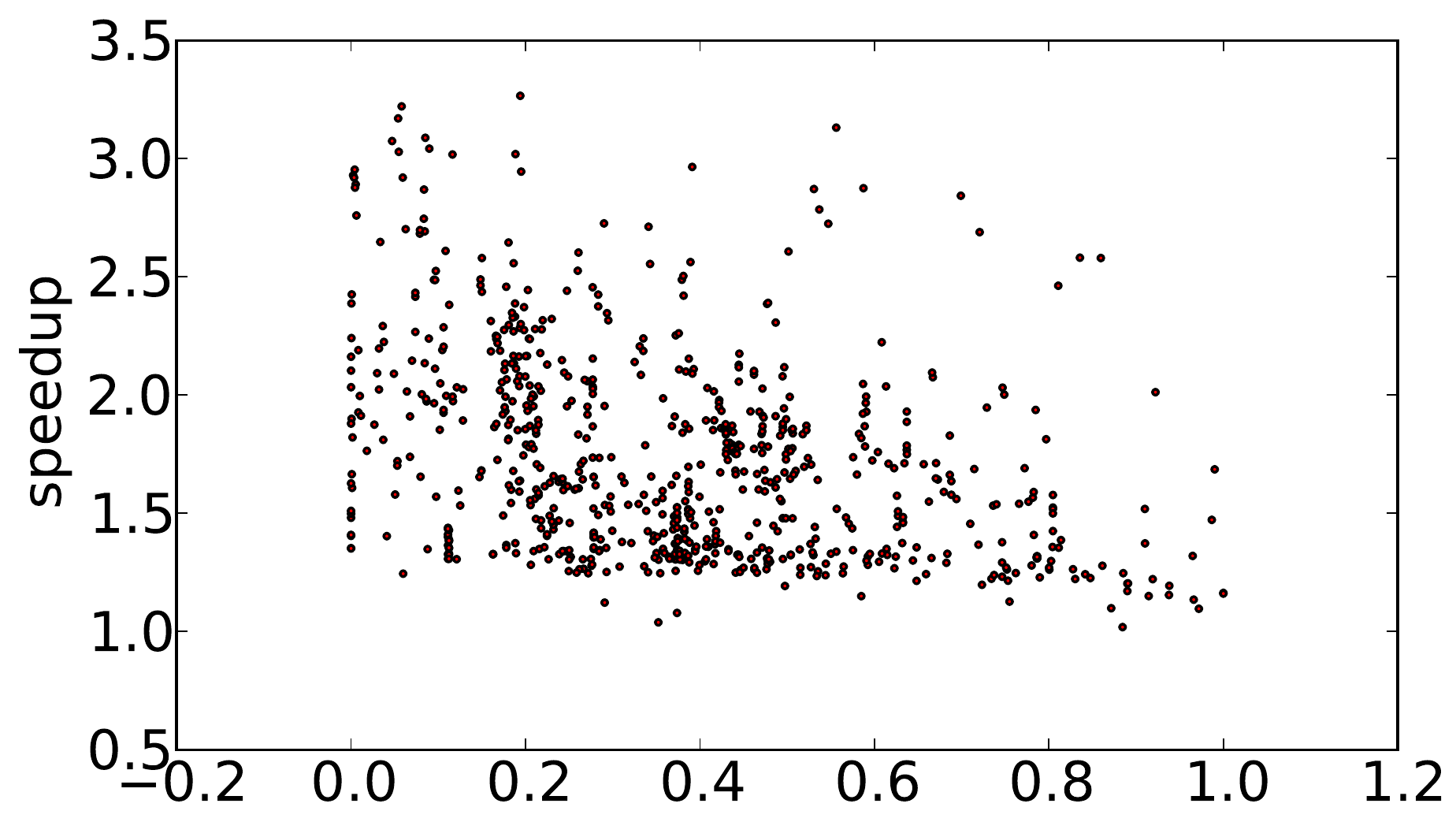}}	
	\subfigure[nnz\_var]{\label{fig:nnz-var-bar}\includegraphics[width=0.45\textwidth]
		{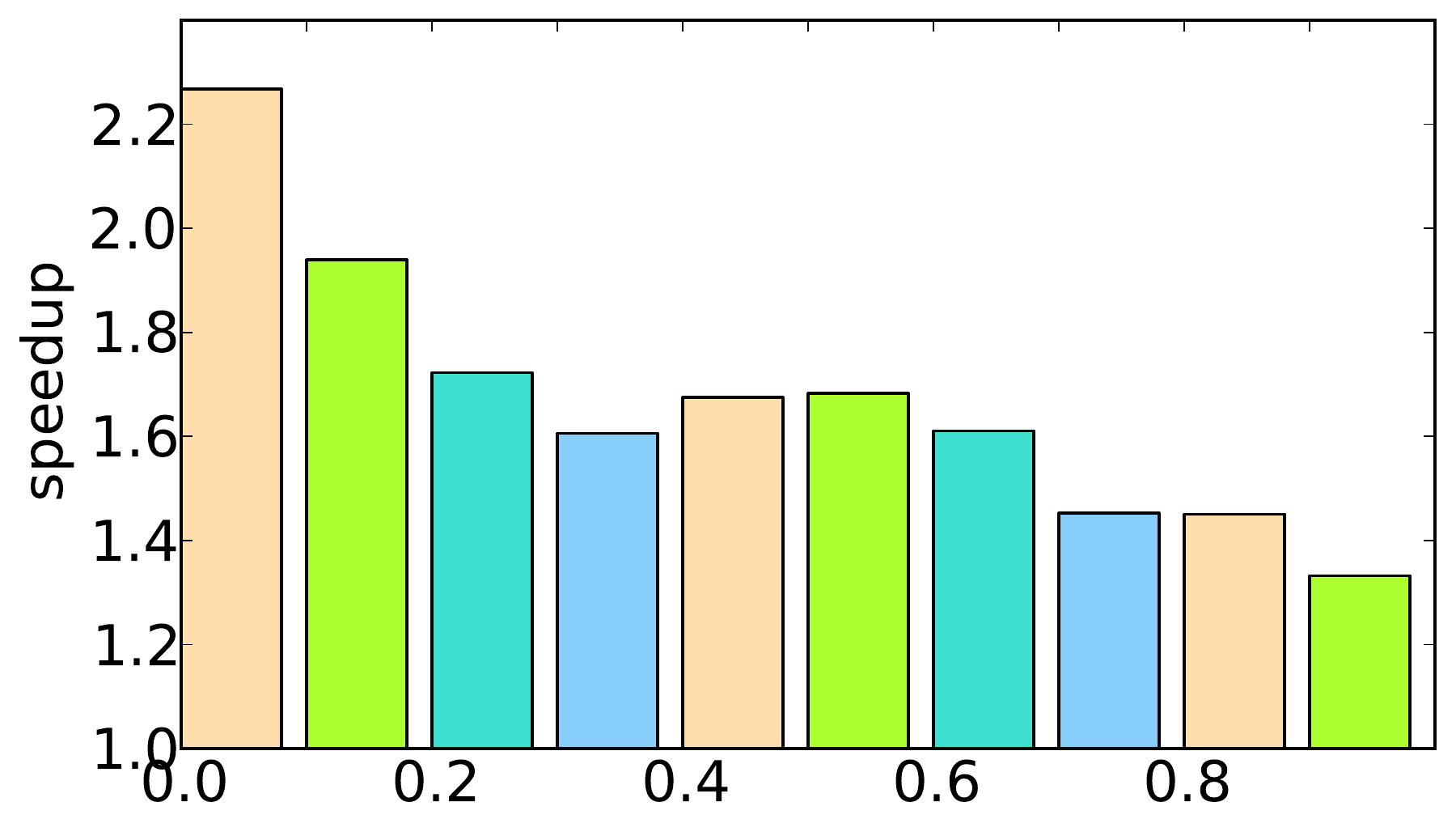}}
	\caption{The correspondence between the three identified factors and speedup of SpMV.
		The y-axis in subfigure (b), (d) and (f) is interval average values of
		speedup.
		In subfigure (e) and (f), the x-axis represents the value of \texttt{nnz\_var} after normalization processing. }
	\label{fig:factors}
	
\end{figure*}


\subsection{SpMV Scalability Factors}
Based on the data obtained from executing SpMV on datasets,
we draw scatter plots between each impacting factor and the speedup, which are shown in Figure~\ref{fig:factors}.
It is clear that the speedup generally shows a gradual decline trend when the nonzero allocation across threads becomes more unbalanced,
the L2\_DCMR increases from one thread to four threads, or the nonzero variance of sparse matrices go larger.

The three bar charts in Figure~\ref{fig:factors} show the statistical results
of integral histogram of the speedup results, which is consistent with the results in the left part of Figure~\ref{fig:factors}.
There are also some cases that do not meet our expectations.
For example, Figure~\ref{fig:L2-cache-bar} shows that the speedup even decreases
when \texttt{L2\_DCMR\_change} is less than 0.
We argue that it is a comprehensive product of multiple impacting factors, which needs further investigation.
In this following, we will analyze how each factor has an impact on the SpMV scalability.

\cparagraph{The balance of the nonzero allocation}
When running the conventional SpMV code in the CSR format, the nonzero allocation across threads depends on the sparse matrix structure.
As shown in Figure~\ref{fig:job-var-bar}, when \texttt{job\_var} is greater than 0.45, which means that the nonzeros are clustered within some rows to be dispatched to a specific thread,
load imbalance will occur and
this thread will take substantially more time than the other threads.
Thus, the unbalanced nonzero allocation among threads will put a limit on the achieved speedup,
because the SpMV performance is determined by the slowest thread.

Taking \texttt{exdata\_1} in Table~\ref{tbl:mtx_structure} for instance, the second thread will consume more than 99\% of the nonzeros when using 4 threads, and thus the achieved speedup stays around 1.02x in such a case.

\begin{table*}[!t]
	\centering
	\caption{The concise description of four representative matrices.}
	\label{tbl:mtx_structure}
	\scalebox{0.7}{
		\begin{tabular}{p{2cm}p{1.5cm}p{3.5cm}p{1.5cm}p{3.5cm}p{3cm}}
			\toprule
			\rowcolor{white}\textbf{matrix} &\textbf{job\_var}         & \textbf{L2\_DCMR\_change}        & \textbf{nnz\_var}   & \textbf{sparsity structure}   &    \textbf{speedup}         \\
			\midrule
			\textbf{exdata\_1}  & 0.992   & 0.000			& 649.627 	&  \begin{minipage}{\textwidth}
				\includegraphics[width=35mm, height=25mm]{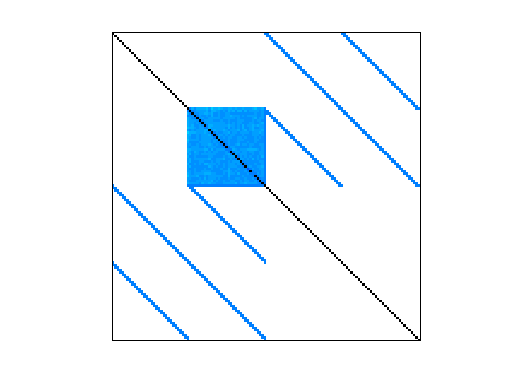}
			\end{minipage}	
			& 1.018x	\\
			
			\textbf{conf5\_4-8x8-20}  & 0.250 	  & 0.056		& 0.000 &  \begin{minipage}{\textwidth}
				\includegraphics[width=35mm, height=25mm]{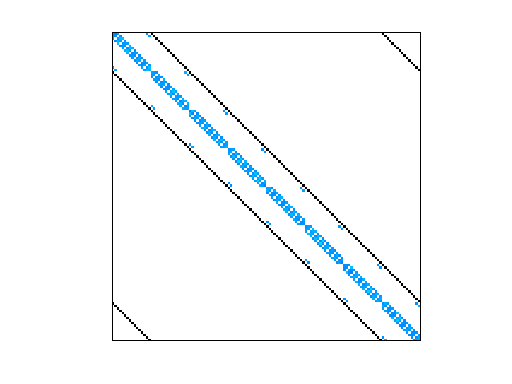}
			\end{minipage}	
			& 1.351x	\\			

			\textbf{debr}  & 0.250 	  & -0.001			& 0.003 	&  \begin{minipage}{\textwidth}
				\includegraphics[width=35mm, height=25mm]{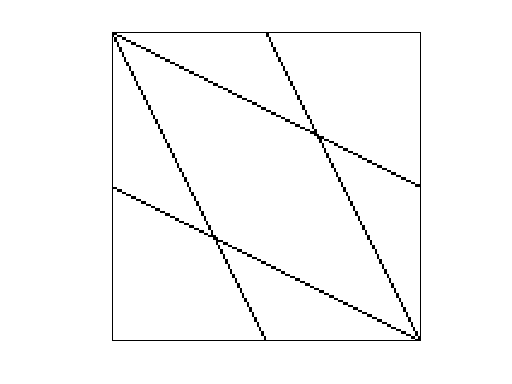}
			\end{minipage}	
			& 2.241x	\\

			\textbf{appu}  & 0.252 	  & -0.001		& 36.494  	&  \begin{minipage}{\textwidth}
				\includegraphics[width=35mm, height=25mm]{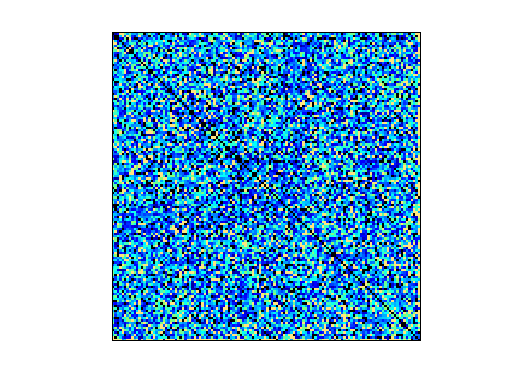}
			\end{minipage}	
			& 1.479x
			
			\\ \bottomrule
		\end{tabular}}

	\end{table*}

\cparagraph{The shared L2 data cache\label{L2-analysis}}
Leveraging shared resources on multi-core architectures improves the utilization of a hardware component and
can improve overall system throughput.
On the one hand, such a design as cache sharing can lead to
positive interference, i.e.,  one thread brings data into the shared cache which is accessed by other threads~\cite{DBLP:conf/ispass/EyermanBE12}.
The \texttt{debr} in Table~\ref{tbl:mtx_structure} gives an intuitive example for the benefits of shared memory.
Recalling the SpMV algorithm, $\mathbf{y}=\mathbf{Ax}$, where the dense vector $\mathbf{x}$ is the data structure to be reused across threads.
This occurs because different threads deal with distinct matrix rows of $\mathbf{A}$, while $\mathbf{x}$ is shared by all threads.
When running SpMV on \texttt{debr} with 4 threads on \texttt{FT-2000+}, with the L2 cache sharing within a core-group, threads[1, 3] and threads[2, 4] can share the dense vector $\mathbf{x}$ so as to increase the data reuse and improve the performance of multi-threaded SpMV.

On the other hand, cache sharing can have a negative impact on the per-thread performance from the perspective of resource competition.
The L2 cache sharing on \texttt{FT-2000+} may cause threads to evict data of other threads when running SpMV,
which means that the `victim' threads will experience more cache misses than their isolated execution.
And we find that the degree of the negative impact from cache sharing is related to the average nonzeros per row (\texttt{nnz\_avg}).
In general, a larger \texttt{nnz\_avg} leads to more competitions.
We argue that this is because \texttt{nnz\_avg} represents the need for dense vector $\mathbf{x}$ per row when running SpMV,
which means that the data evicting increases as \texttt{nnz\_avg} goes up.
As shown in Figure~\ref{fig:L2-cache}, as \texttt{L2\_DCMR} increases for most matrices,
we note a corresponding decrease in speedup.

To summarize, 
we note that the impact of cache sharing on SpMV relates to specific input matrices and their structures.
In Table~\ref{tbl:mtx_structure}, the SpMV gains a much larger speedup on \texttt{debr} (2.241x) than on \texttt{conf5\_4-8x8-20} (1.351x) with 4 threads. On the one hand, the data reuse that benefits from the distribution of nonzeros makes contributions;
on the other hand, the average nonzeros per row of \texttt{conf5\_4-8x8-20} is larger than its counterpart (39 vs 4),
which means that runnig SpMV on \texttt{conf5\_4-8x8-20} generates
more contention with shared L2 cache. These two reasons both lead to a higher increase on \texttt{L2\_DCMR} of \texttt{conf5\_4-8x8-20} than \texttt{debr} from one thread to four threads.



\cparagraph{The nonzero variance across rows}
The utilization of the dense vector $\mathbf{x}$ has a significant impact on the SpMV scalability.
However, to obtain the correlation of nonzero distribution row by row is time-consuming for large-scale sparse matrices.
As a result, we choose the nonzero variance across rows instead.
This metric can reflect the regularity of input matrices and capture how the dense vector $\mathbf{x}$ will be reused.

Note that the speedup is calculated by dividing the single-thread execution time by the that of multiple threads,
and the latter depends on the thread that spends the most time.
Thus, an even distribution of the SpMV execution across threads typically leads to satisfactory SpMV scalability.
However, we observe that the balanced nonzero allocation across rows does not necessarily lead to a large speedup like matrix \texttt{debr} listed in Table~\ref{tbl:mtx_structure}.
This is because the different nonzeros distribution across rows (and threads)
equally has an impact on the execution time.
For \texttt{debr}, despite the fact that nonzeros are evenly allocated across threads, the large \texttt{nnz\_var} results in different reuse of vector $\mathbf{x}$, and leads to different execution behaviours across threads and
an unsatisfactory speedup.
As shown in Figure~\ref{fig:nnz-var-bar}, matrices with smaller \texttt{nnz\_var} tend to bring a larger speedup.
This can be equally explained that a smaller nonzero variance across rows can ensure that the workloads can be more evenly distributed and better
exploit the locality of vector $\mathbf{x}$.

%% file: spmv_advise.tex
\subsection{Potential Optimizations}

\begin{figure}[!t]
	\centering
	\includegraphics[width=0.5\textwidth]{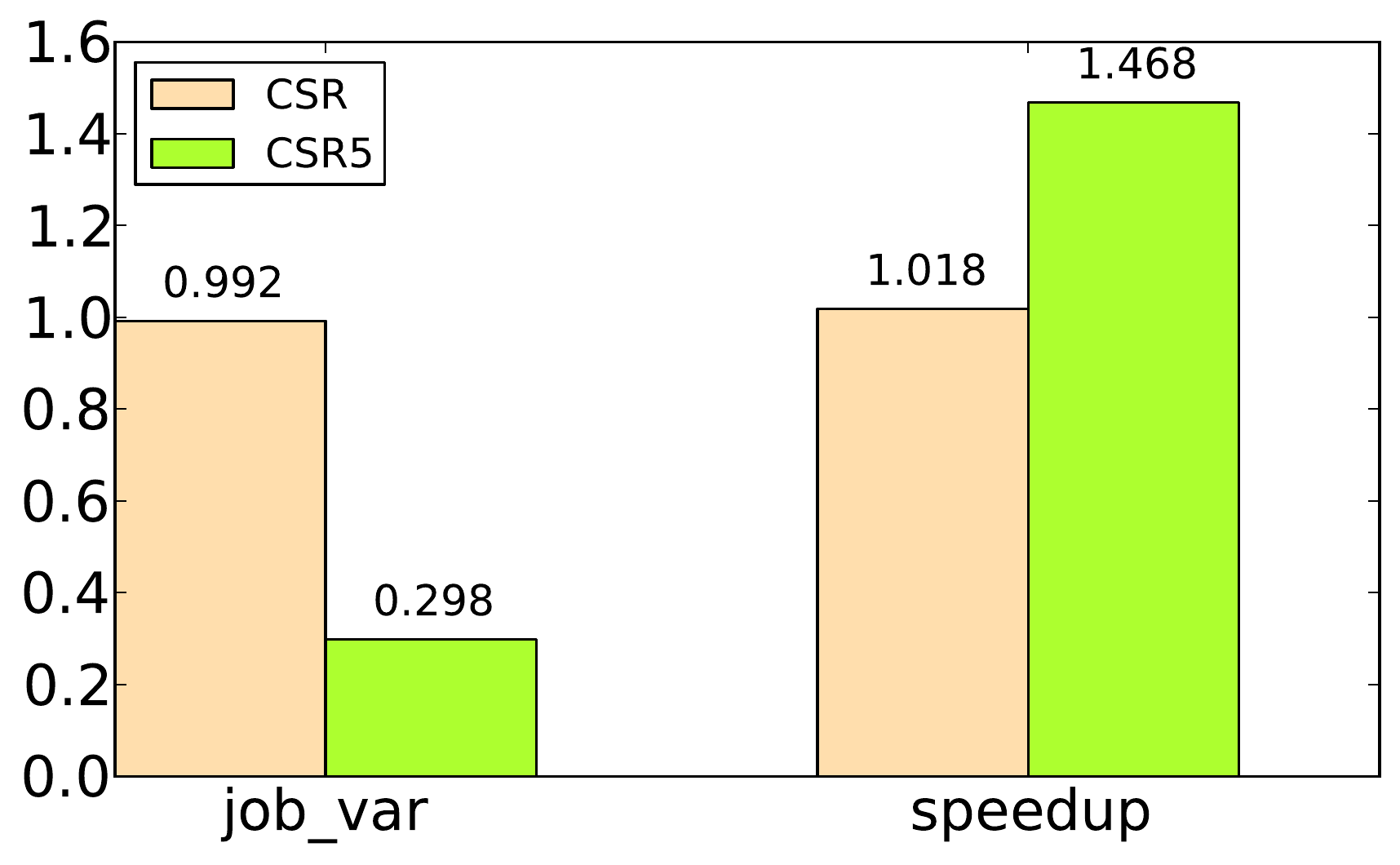}
	\caption{The comparison of \texttt{job\_var} and speedup (normalized to that of a single thread) of SpMV in different storage formats. The input matrix is \texttt{exdata\_1}.}
	\label{fig:CSR5vsCSR}
	
\end{figure}

\subsubsection{Using storage formats with balanced nonzero allocation}
The load imbalance is mainly related to the adopted CSR format and the thread scheduling policy.
In most cases, we use the static scheduling policy because the overhead of thread communication with dynamic scheduling is nonnegligible.
To overcome the issue of load imbalance, we choose to use storage formats that divide nonzeros equally among threads.
The CSR5 format is selected because it is designed to solve the load imbalance in CSR-based SpMV, and its data structure is shown in Section~\ref{sec:format}.

We choose matrices whose scalability suffers from load imbalance by its  \texttt{job\_var} value ($\geq$ 0.45), and then run CSR5-based SpMV on the matrices. The results show CSR5 achieves an average improvement of speedup from 1.632x to 2.023x.
Figure~\ref{fig:CSR5vsCSR} shows the performance result on \texttt{exdata\_1}.
Compared with the CSR format, load imbalance is
significantly mitigated by running the CSR5-based SpMV
with \texttt{job\_var} decreasing from 0.992 to 0.298.
As a consequence, the speedup gains an improvement from 1.018x to 1.468x.
CSR5 performs better
because the nonzeros are divided and organized in small tiles
instead of the row manner.
Therefore, when dealing with irregular matrices, despite that the rows with a large number of nonzeros may be concentrated like \texttt{exdata\_1},
they will not be assigned to the same thread.
Thus, the workloads can be dispatched in a much more even manner across threads and improve the scalability of SpMV.

\begin{figure}[!t]
	\centering
	\includegraphics[width=0.5\textwidth]{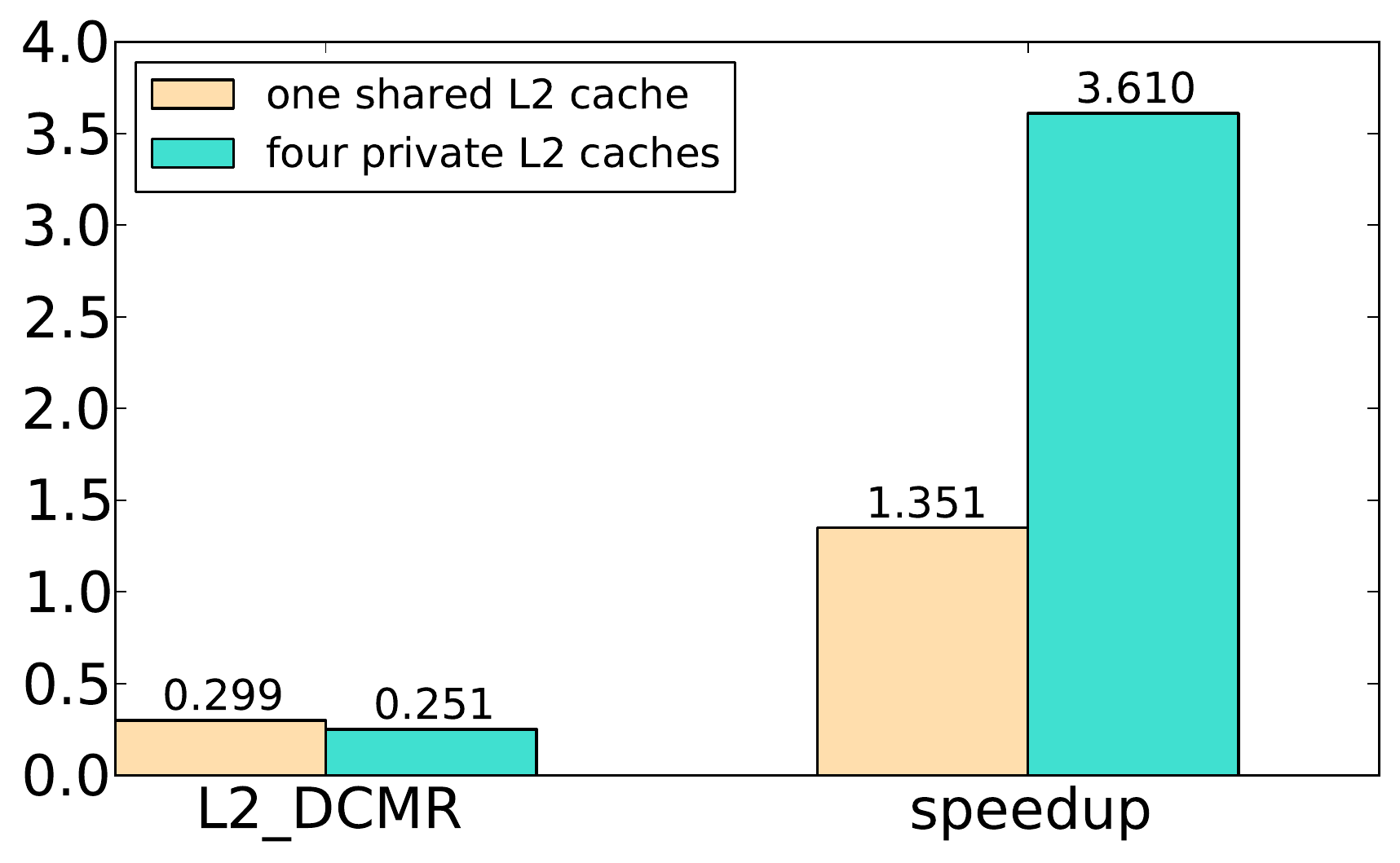}
	\caption{The SpMV scalability improvement benefited from our optimizations on \texttt{conf5\_4-8x8-20}.}
	\label{fig:one-vs-muti}
	
\end{figure}

\subsubsection{Avoiding the contention from shared memory resources}
Based on the analysis in Section~\ref{L2-analysis},
we know that the sharing L2 cache of \texttt{FT-2000+} has a great impact on the scalability of SpMV.
Under most circumstances, the sharing cache causes more contention, which leads to a speedup decline.
To alleviate the pressure from cache sharing,
we bind threads to multiple cores that are located in different \texttt{core-groups} (Section~\ref{sec:setup}).
In this way, we can ensure that each thread occupies a complete L2 cache without data interference from other threads.

When running SpMV on all the matrices in the private-L2 mode, we can achieve a considerable average speedup of 3.40x on 4 threads, compared with 1.93x on one core-group (Table~\ref{tbl:average_speedup}).
As can be seen from Figure~\ref{fig:one-vs-muti}, the speedup with private L2 caches significantly outperforms its counterpart of sharing an L2 data cache on \texttt{conf5\_4-8x8-20}, with a speedup increasing from 1.35x to 3.61x.
This is because using private L2 caches can effectively reduce the L2 cache miss from 30\% to 25\%.
But this approach of using a private L2 data cache
will not bring a performance increase
for all matrices.
Taking another matrix \texttt{asia\_osm} for example,
the speedup only increases by 2.6\% from 3.170x to 3.254x with private L2 caches.
We reckon that the average nonzeros per row of this matrix is less than 3, so that the shared L2 cache can meet with their memory accessing need.

%

\begin{figure}[!t]
	\centering
	\includegraphics[width=0.5\textwidth]{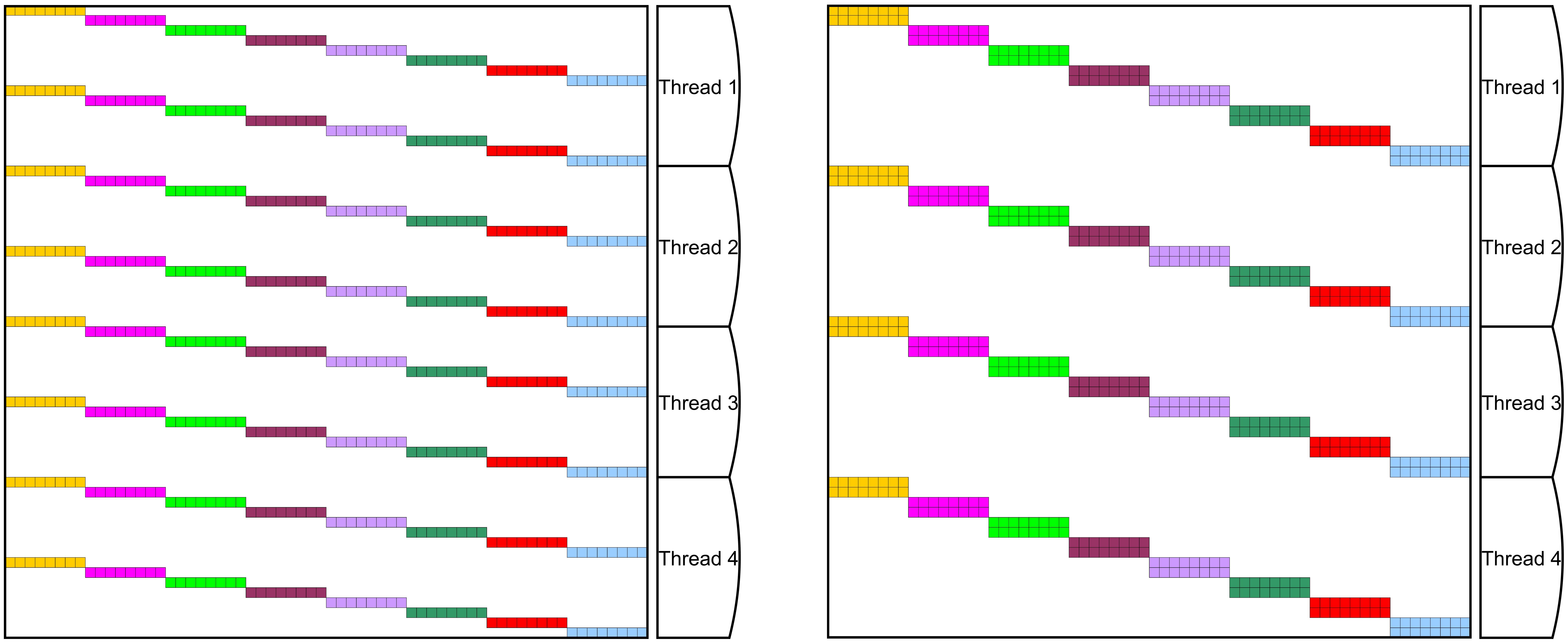}
	\caption{The synthesized sparse matrix with poor locality utilization of the vector $\mathbf{x}$ (left) and the corresponding matrix in ideal locality-aware SpMV storage format (right).}
	\label{fig:locality}
	
\end{figure}

\subsubsection{Exploiting locality-aware SpMV storage formats}
Based on the aforementioned analysis, we know that merely achieving balanced nonzero allocation is insufficient, and the locality of $\mathbf{x}$ in SpMV needs to be exploited to achieve better scalability.
Here is our idea to design a novel storage format that can make good use of the locality:
we bring together the rows with a similar nonzero distribution, so that the vector $\mathbf{x}$ can be reused.

To explore the feasibility of designing the locality-aware SpMV storage format, we generate a series of matrices of different sizes as shown in Figure~\ref{fig:locality}.
This synthesized matrix has a poor locality of vector $\mathbf{x}$ when running SpMV.
And we then transform such matrices to locality-friendly forms by partial reordering.
Table~\ref{tbl:locality_compare} shows the result of running CSR-based SpMV on a specific pair of matrices on \texttt{FT-2000+}.
Both single-threaded and multi-threaded performance gain significant improvement.
Particularly, the 64-thread performance improves by 71.7\% from 15.907 Gflops to 27.306 Gflops.
At the same time, better scalability of SpMV is achieved from 37.96x to 46.68x.


\vspace{2mm}
To conclude, we introduce several potential optimizations inspired by the scalability results,
but these are not one-fit-all solutions.
This is because there is an overhead for format conversion, and
using multiple private L2 caches waste extra memory resources.
For future work, we will extract a detailed profile of a given sparse matrix before performing the SpMV
computation.
Hopefully, these features will indicate the number and distribution of nonzeros.
Based on this information, we can decide whether to apply these optimizations or not.
Besides, we will try to find an accurate and efficient matrix reordering that can be applied to design the locality-aware SpMV storage format.

\begin{table*}[!t]
	\centering
	\caption{The performance and scalability of SpMV by exploiting the locality of $\mathbf{x}$. The rows of the matrix is set to 64*6400, with
	the average nonzeros per rows being 4.}
	\label{tbl:locality_compare}
	\scalebox{0.80}{
		\begin{tabular}{@{}lllll@{}}
			\toprule
			\rowcolor{white} &\textbf{single-thread Perf.}        & \textbf{64-thread Perf.}        & \textbf{speedup}           \\
			\midrule
			\rowcolor{Gray}\textbf{synthesized matrix}  & 0.419 Gflops	  & 15.907 Gflops		&37.96x	 \\		
			
			\textbf{transformed matrix}  & 0.585 Gflops 	  & 27.306 Gflops	&46.68x		\\ \bottomrule
		\end{tabular}}
		
	\end{table*}

%% file: spmv_relatedwork.tex
\section{Related Work}

Substantial previous work has been conducted to study the SpMV performance on parallel
systems~\cite{DBLP:journals/ijhpca/Mellor-CrummeyG04, DBLP:conf/sc/PinarH99, DBLP:journals/pc/WilliamsOVSYD09,IJPP/Chen19}.
Mellor-Crummey et. al use a loop transformation to improve the performance of SpMV on multiple parallel processors,
and this optimization is aimed at the matrices that arise in SAGE~\cite{DBLP:journals/ijhpca/Mellor-CrummeyG04}.
Pinar et. al propose alternative data structures, along with reordering algorithms to reduce the number of memory indirections when running SpMV on a Sun Enterprise 3000~\cite{DBLP:conf/sc/PinarH99}.
Williams et. al apply several optimization strategies especially effective for the multicore environment to SpMV on four multicore platforms.
These works have a significant effect on improving the performance of parallel SpMV~\cite{DBLP:journals/pc/WilliamsOVSYD09}.
However, very few works focus on its scalability performance on many-core architectures.
Our work fills this gap by providing an in-depth scalability analysis on \texttt{FT-2000+}.
The obtained insights will facilitate us to design more efficient parallel HPC software and hardware in the future.

Efforts have been made in designing new storage formats for various parallel processor architectures
including SIMD CPUs and SIMT
GPUs~\cite{DBLP:conf/sc/BellG09,DBLP:conf/iccS/MaggioniB13,DBLP:conf/ics/0002V15,DBLP:conf/hipeac/MonakovLA10,DBLP:journals/siamsc/KreutzerHWFB14}.
Bell et. al use standard CUDA idioms to implement several SpMV kernels which can exploit fine-grained parallelism to
effectively utilize the computational resources of GPUs,
including SIMD-friendly ELL, the most popular general-purpose CSR and hybrid ELL/COO format that exploits the advantages of both~\cite{DBLP:conf/sc/BellG09}.
The CSR5 proposed by Liu et. al is efficient both for regular matrices and for
irregular matrices and is also used in our work to address the issue of unbalanced loads~\cite{DBLP:conf/ics/0002V15}.
Maggioni et. al propose the design of an architecture-aware technique for improving the performance of the SpMV on GPU,
and based on a variation of the sliced ELL sparse format, they present a warp-oriented ELL format that is suited for regular matrices~\cite{DBLP:conf/iccS/MaggioniB13}.
The SELL-C-$\sigma$ format is designed by Kreutzer et. al, and this SIMD-friendly data format is well-suited for a variety of hardware platforms (Intel Sandy Bridge, Intel Xeon Phi, and Nvidia Tesla K20)~\cite{DBLP:journals/siamsc/KreutzerHWFB14}.
These sparse matrix formats aim to address the issue of unbalanced load and increase SpMV parallelism, but they fail to take advantage of the locality of vector $\mathbf{x}$.
Our work attempts to answer this question by providing comprehensive analysis and new insights.

A large number of works have analyzed the sources of poor scalability in various parallel applications, rather than SpMV~\cite{DBLP:conf/ispass/DiamondBMKKB11,DBLP:conf/ics/LiuLS08,DBLP:conf/iiswc/AlamBKRV06}.
Alam et. al propose an appropriate selection of MPI task and memory placement schemes to improve performance for key scientific calculations on multi-core AMD Opteron processors~\cite{DBLP:conf/iiswc/AlamBKRV06}.
Liu et. al introduce the notion of
memory access intensity to facilitate quantitative analysis of
program’s memory behavior on multicores~\cite{DBLP:conf/ics/LiuLS08}.
For the work of Diamond et. al, it not only examines traditional unicore metrics and but also presents an in-depth study of performance bottlenecks
originating in multicore-based systems. Besides, it introduces a source-code optimization called loop microfission to alleviate
multicore-related performance bottlenecks~\cite{DBLP:conf/ispass/DiamondBMKKB11}.
Bhattacharjee et. al~\cite{DBLP:conf/isca/BhattacharjeeM09} predict critical threads, or threads that suffer from imbalance.
They tend to offer more resources to critical threads so that they run faster.
Most of these related works focus on the traditional \texttt{x86} multi-core architectures, and very few works are towards the ARMv8-based many-cores or the SpMV kernel, which is rather promising for the future of the HPC domain.

Numerous performance analysis tools have been proposed, including \textit{CounterMiner} and \textit{HPCTOOLKIT}~\cite{DBLP:conf/micro/LvSLW0Q18,DBLP:journals/concurrency/AdhiantoBFKMMT10}.
By using data mining and machine learning techniques, \texttt{CounterMiner} enables the measurement and understanding of big performance data~\cite{DBLP:conf/micro/LvSLW0Q18}.
\texttt{HPCTOOLKIT} can pinpoint and quantify scalability bottlenecks of fully optimized parallel programs.
Based on statistical sampling, this tool can introduce with a very small measurement overhead during performance measurement~\cite{DBLP:journals/concurrency/AdhiantoBFKMMT10}.
Different from these performance tools, our regression-tree based approach uses both hardware counters (dynamic features) and input matrix features (static features),
thus brings a comprehensive understanding of the scalability behaviours.

Other researchers have used performance counters to identify multicore bottlenecks and optimize applications~\cite{DBLP:conf/ics/LiuLS08},
but no quantitative analysis is performed in those studies. Our work not only conducts detailed scalability analysis, but also is the first
attempt in applying machine learning techniques to find the impact factors of SpMV scalability on \texttt{FT-2000+}.

Machine learning has quickly emerged as a powerful design methodology for systems modeling and
optimization~\cite{mlcpieee}. Prior works have demonstrated the success of applying machine learning for a wide range
of tasks, including modeling code
optimization~\cite{wang2014integrating,Tournavitis:2009:THA:1542476.1542496,Wang:2009:MPM:1504176.1504189,wang2010partitioning,grewe2013portable,wang2013using,DBLP:journals/taco/WangGO14,
ogilvie2014fast,cummins2017end,ogilvie2017minimizing,spmv,ipdpsz18,ijpp18,yuan2019using,tecs19}, task
scheduling~\cite{grewe2011workload,emani2013smart,grewe2013opencl,marco2017improving}, processor resource
allocation~\cite{wen2014smart}, and many others~\cite{ren2017optimise,Ren:2018:PNW:3281411.3281422}. In this work, we
employ machine learning techniques to develop an automatic and portable approach to characterize the scalability of
SpMV on an emerging many-core architecture. We stress that this work does not seek to advance machine learning
algorithms; instead, it explores and applies a well-established modeling method to tackle the optimization problem for
an important class of applications.

%% file: spmv_conclusion.tex
\section{Conclusion}
This paper has presented an empirical study of SpMV scalability on an emerging ARMv8-based many-core architecture, Phytium \texttt{FT-2000+}. 
We conduct an overall evaluation about the scalability of SpMV on \texttt{FT-2000+}. 
We develop a machine learning based model to help find the main factors that lead to the flat scalability: 
unbalanced nonzero allocation, shared L2 cache contention and nonzero variance per row. 
We use a statistical method to find the relations between factors and the speedup of SpMV as a verification of our model. 
We select representative matrices to explain how these factors give a limit to the scalability of SpMV on \texttt{FT-2000+} in an essential way not remain it in ``black box". 
Along the line, we give potential optimizations 
for mitigating these scalability bottlenecks on SpMV. 
Our experimental results show that our optimization can effectively improve the scalability of specific matrices.

%% file: aspmv.bbl
\begin{thebibliography}{53}
\providecommand{\natexlab}[1]{#1}
\providecommand{\url}[1]{{#1}}
\providecommand{\urlprefix}{URL }
\expandafter\ifx\csname urlstyle\endcsname\relax
  \providecommand{\doi}[1]{DOI~\discretionary{}{}{}#1}\else
  \providecommand{\doi}{DOI~\discretionary{}{}{}\begingroup
  \urlstyle{rm}\Url}\fi
\providecommand{\eprint}[2][]{\url{#2}}

\bibitem[{ft2(2017)}]{ft2000plus}
 (2017) FT-2000 Plus. Phytium Technology Co. Ltd.,
  \url{http://tech.sina.com.cn/d/2017-10-16/doc-ifymvuyt0962449.shtml}

\bibitem[{Adhianto et~al.(2010)Adhianto, Banerjee, Fagan, Krentel, Marin,
  Mellor{-}Crummey, and Tallent}]{DBLP:journals/concurrency/AdhiantoBFKMMT10}
Adhianto L, Banerjee S, Fagan MW, Krentel M, Marin G, Mellor{-}Crummey JM,
  Tallent NR (2010) {HPCTOOLKIT:} tools for performance analysis of optimized
  parallel programs. Concurrency and Computation: Practice and Experience pp
  685--701

\bibitem[{Alam et~al.(2006)Alam, Barrett, Kuehn, Roth, and
  Vetter}]{DBLP:conf/iiswc/AlamBKRV06}
Alam SR, Barrett RF, Kuehn JA, Roth PC, Vetter JS (2006) Characterization of
  scientific workloads on systems with multi-core processors. In: Proceedings
  of the 2006 {IEEE} International Symposium on Workload Characterization,
  {IISWC} 2006, October 25-27, 2006, San Jose, California, {USA}, pp 225--236

\bibitem[{Bell and Garland(2009)}]{DBLP:conf/sc/BellG09}
Bell N, Garland M (2009) Implementing sparse matrix-vector multiplication on
  throughput-oriented processors. In: SC

\bibitem[{Benatia et~al.(2016)Benatia, Ji, Wang, and
  Shi}]{DBLP:conf/icpp/BenatiaJWS16}
Benatia A, Ji W, Wang Y, Shi F (2016) Sparse matrix format selection with
  multiclass {SVM} for spmv on {GPU}. In: 45th International Conference on
  Parallel Processing, {ICPP} 2016, Philadelphia, PA, USA, August 16-19, 2016,
  pp 496--505

\bibitem[{Bhattacharjee and Martonosi(2009)}]{DBLP:conf/isca/BhattacharjeeM09}
Bhattacharjee A, Martonosi M (2009) Thread criticality predictors for dynamic
  performance, power, and resource management in chip multiprocessors. In: 36th
  International Symposium on Computer Architecture {(ISCA} 2009), June 20-24,
  2009, Austin, TX, {USA}, pp 290--301

\bibitem[{Chen et~al.(2019)Chen, Fang, Chen, Xu, and Wang}]{IJPP/Chen19}
Chen D, Fang J, Chen S, Xu C, Wang Z (2019) Optimizing sparse matrix-vector
  multiplications on an armv8-based many-core architecture. International
  Journal of Parallel Programming 47(3):418--432

\bibitem[{Chen et~al.(2018)}]{spmv}
Chen S, et~al. (2018) Adaptive optimization of sparse matrix-vector
  multiplication on emerging many-core architectures. In: HPCC '18

\bibitem[{Cummins et~al.(2017)Cummins, Petoumenos, Wang, and
  Leather}]{cummins2017end}
Cummins C, Petoumenos P, Wang Z, Leather H (2017) End-to-end deep learning of
  optimization heuristics. In: PACT

\bibitem[{Davis and Hu(2011)}]{DBLP:journals/toms/DavisH11}
Davis TA, Hu Y (2011) The university of florida sparse matrix collection. {ACM}
  Trans Math Softw

\bibitem[{Diamond et~al.(2011)Diamond, Burtscher, McCalpin, Kim, Keckler, and
  Browne}]{DBLP:conf/ispass/DiamondBMKKB11}
Diamond JR, Burtscher M, McCalpin JD, Kim B, Keckler SW, Browne JC (2011)
  Evaluation and optimization of multicore performance bottlenecks in
  supercomputing applications. In: {IEEE} International Symposium on
  Performance Analysis of Systems and Software, {ISPASS} 2011, 10-12 April,
  2011, Austin, TX, {USA}, pp 32--43

\bibitem[{Emani et~al.(2013)Emani, Wang, and O'Boyle}]{emani2013smart}
Emani MK, Wang Z, O'Boyle MFP (2013) Smart, adaptive mapping of parallelism in
  the presence of external workload. In: CGO

\bibitem[{Eyerman et~al.(2012)Eyerman, Bois, and
  Eeckhout}]{DBLP:conf/ispass/EyermanBE12}
Eyerman S, Bois KD, Eeckhout L (2012) Speedup stacks: Identifying scaling
  bottlenecks in multi-threaded applications. In: 2012 {IEEE} International
  Symposium on Performance Analysis of Systems {\&} Software, New Brunswick,
  NJ, USA, April 1-3, 2012, pp 145--155

\bibitem[{Grewe et~al.(2011)Grewe, Wang, and O'Boyle}]{grewe2011workload}
Grewe D, Wang Z, O'Boyle MFP (2011) A workload-aware mapping approach for
  data-parallel programs. In: HiPEAC

\bibitem[{Grewe et~al.(2013{\natexlab{a}})Grewe, Wang, and
  O'Boyle}]{grewe2013portable}
Grewe D, Wang Z, O'Boyle MFP (2013{\natexlab{a}}) Portable mapping of data
  parallel programs to opencl for heterogeneous systems. In: CGO

\bibitem[{Grewe et~al.(2013{\natexlab{b}})}]{grewe2013opencl}
Grewe D, et~al. (2013{\natexlab{b}}) Opencl task partitioning in the presence
  of gpu contention. In: LCPC

\bibitem[{Gupta et~al.(2012)Gupta, Kim, and Schwan}]{Gupta2012Evaluating}
Gupta V, Kim H, Schwan K (2012) Evaluating scalability of multi-threaded
  applications on a many-core platform. Georgia Institute of Technology

\bibitem[{Kincaid et~al.(1989)Kincaid, D.R, T.C, and Young}]{osti_7093021}
Kincaid, DR, TC, Young (1989) Itpackv 2d user's guide. Tech. rep., Center for
  Numerical Analysis, Texas Univ., Austin, TX (USA)

\bibitem[{Kreutzer et~al.(2014)Kreutzer, Hager, Wellein, Fehske, and
  Bishop}]{DBLP:journals/siamsc/KreutzerHWFB14}
Kreutzer M, Hager G, Wellein G, Fehske H, Bishop AR (2014) A unified sparse
  matrix data format for efficient general sparse matrix-vector multiplication
  on modern processors with wide {SIMD} units. {SIAM} J Scientific Computing

\bibitem[{Laurenzano et~al.(2016)Laurenzano, Tiwari, Cauble{-}Chantrenne,
  Jundt, Jr., Campbell, and Carrington}]{DBLP:conf/ispass/LaurenzanoTCJWC16}
Laurenzano MA, Tiwari A, Cauble{-}Chantrenne A, Jundt A, Jr WAW, Campbell RL,
  Carrington L (2016) Characterization and bottleneck analysis of a 64-bit
  armv8 platform. In: ISPASS

\bibitem[{Lindong et~al.(2018)}]{ijpp18}
Lindong C, et~al. (2018) Optimizing sparse matrix-vector multiplications on an
  armv8-based many-core architecture. International Journal of Parallel
  Programming

\bibitem[{Liu et~al.(2018)Liu, He, Liu, and Tan}]{DBLP:conf/ppopp/LiuHLT18}
Liu J, He X, Liu W, Tan G (2018) Register-based implementation of the sparse
  general matrix-matrix multiplication on gpus. In: PPoPP

\bibitem[{Liu et~al.(2008)Liu, Li, and Sameh}]{DBLP:conf/ics/LiuLS08}
Liu L, Li Z, Sameh AH (2008) Analyzing memory access intensity in parallel
  programs on multicore. In: Proceedings of the 22nd Annual International
  Conference on Supercomputing, {ICS} 2008, Island of Kos, Greece, June 7-12,
  2008, pp 359--367

\bibitem[{Liu and Vinter(2015{\natexlab{a}})}]{DBLP:conf/ics/0002V15}
Liu W, Vinter B (2015{\natexlab{a}}) {CSR5:} an efficient storage format for
  cross-platform sparse matrix-vector multiplication. In: ICS

\bibitem[{Liu and Vinter(2015{\natexlab{b}})}]{DBLP:journals/pc/0002V15}
Liu W, Vinter B (2015{\natexlab{b}}) Speculative segmented sum for sparse
  matrix-vector multiplication on heterogeneous processors. Parallel Computing
  49:179--193

\bibitem[{Lv et~al.(2018)Lv, Sun, Luo, Wang, Yu, and
  Qian}]{DBLP:conf/micro/LvSLW0Q18}
Lv Y, Sun B, Luo Q, Wang J, Yu Z, Qian X (2018) Counterminer: Mining big
  performance data from hardware counters. In: 51st Annual {IEEE/ACM}
  International Symposium on Microarchitecture, {MICRO} 2018, Fukuoka, Japan,
  October 20-24, 2018, pp 613--626

\bibitem[{Maggioni and Berger{-}Wolf(2013)}]{DBLP:conf/iccS/MaggioniB13}
Maggioni M, Berger{-}Wolf TY (2013) An architecture-aware technique for
  optimizing sparse matrix-vector multiplication on gpus. In: ICCS

\bibitem[{Magni et~al.(2013)Magni, Dubach, and
  O'Boyle}]{DBLP:conf/sc/MagniDO13}
Magni A, Dubach C, O'Boyle MFP (2013) A large-scale cross-architecture
  evaluation of thread-coarsening. In: International Conference for High
  Performance Computing, Networking, Storage and Analysis, SC'13, Denver, CO,
  {USA} - November 17 - 21, 2013, pp 11:1--11:11

\bibitem[{Marco et~al.(2017)Marco, Taylor, Porter, and
  Wang}]{marco2017improving}
Marco VS, Taylor B, Porter B, Wang Z (2017) Improving spark application
  throughput via memory aware task co-location: a mixture of experts approach.
  In: Middleware

\bibitem[{Mellor{-}Crummey and
  Garvin(2004)}]{DBLP:journals/ijhpca/Mellor-CrummeyG04}
Mellor{-}Crummey JM, Garvin J (2004) Optimizing sparse matrix - vector product
  computations using unroll and jam. {IJHPCA}

\bibitem[{Monakov et~al.(2010)Monakov, Lokhmotov, and
  Avetisyan}]{DBLP:conf/hipeac/MonakovLA10}
Monakov A, Lokhmotov A, Avetisyan A (2010) Automatically tuning sparse
  matrix-vector multiplication for {GPU} architectures. In: HIPEAC

\bibitem[{Ogilvie et~al.(2014)Ogilvie, Petoumenos, Wang, and
  Leather}]{ogilvie2014fast}
Ogilvie WF, Petoumenos P, Wang Z, Leather H (2014) Fast automatic heuristic
  construction using active learning. In: LCPC

\bibitem[{Ogilvie et~al.(2017)Ogilvie, Petoumenos, Wang, and
  Leather}]{ogilvie2017minimizing}
Ogilvie WF, Petoumenos P, Wang Z, Leather H (2017) Minimizing the cost of
  iterative compilation with active learning. In: CGO

\bibitem[{Pedregosa et~al.(2011)}]{scikitlearn}
Pedregosa F, et~al. (2011) Scikit-learn: Machine learning in python. J Mach
  Learn Res

\bibitem[{Pinar and Heath(1999)}]{DBLP:conf/sc/PinarH99}
Pinar A, Heath MT (1999) Improving performance of sparse matrix-vector
  multiplication. In: SC

\bibitem[{Ren et~al.(2017)Ren, Gao, Wang, and Wang}]{ren2017optimise}
Ren J, Gao L, Wang H, Wang Z (2017) Optimise web browsing on heterogeneous
  mobile platforms: a machine learning based approach. In: INFOCOM

\bibitem[{Ren et~al.(2018)}]{Ren:2018:PNW:3281411.3281422}
Ren J, et~al. (2018) Proteus: Network-aware web browsing on heterogeneous
  mobile systems. In: CoNEXT '18

\bibitem[{{Sanz Marco} et~al.(2019){Sanz Marco}, Taylor, Wang, and
  Elkhatib}]{tecs19}
{Sanz Marco} V, Taylor B, Wang Z, Elkhatib Y (2019) Optimizing deep learning
  inference on embedded systems through adaptive model selection. ACM
  Transactions on Embedded Computing

\bibitem[{Sedaghati et~al.(2015)Sedaghati, Mu, Pouchet, Parthasarathy, and
  Sadayappan}]{DBLP:conf/ics/SedaghatiMPPS15}
Sedaghati N, Mu T, Pouchet L, Parthasarathy S, Sadayappan P (2015) Automatic
  selection of sparse matrix representation on gpus. In: ICS

\bibitem[{Stephens(2016)}]{DBLP:conf/hotchips/Stephens16}
Stephens N (2016) Armv8-a next-generation vector architecture for {HPC}. In:
  2016 {IEEE} Hot Chips 28 Symposium (HCS), pp 1--31

\bibitem[{Terpstra et~al.(2009)Terpstra, Jagode, You, and
  Dongarra}]{DBLP:conf/ptw/TerpstraJYD09}
Terpstra D, Jagode H, You H, Dongarra JJ (2009) Collecting performance data
  with {PAPI-C}. In: Tools for High Performance Computing 2009, pp 157--173

\bibitem[{Tournavitis et~al.(2009)Tournavitis, Wang, Franke, and
  O'Boyle}]{Tournavitis:2009:THA:1542476.1542496}
Tournavitis G, Wang Z, Franke B, O'Boyle MFP (2009) Towards a holistic approach
  to auto-parallelization: Integrating profile-driven parallelism detection and
  machine-learning based mapping. In: PLDI

\bibitem[{Wang and O'Boyle(2018)}]{mlcpieee}
Wang Z, O'Boyle M (2018) Machine learning in compiler optimization. Proc IEEE

\bibitem[{Wang and O'Boyle(2009)}]{Wang:2009:MPM:1504176.1504189}
Wang Z, O'Boyle MF (2009) Mapping parallelism to multi-cores: A machine
  learning based approach. In: PPoPP '09

\bibitem[{Wang and O'Boyle(2010)}]{wang2010partitioning}
Wang Z, O'Boyle MF (2010) Partitioning streaming parallelism for multi-cores: a
  machine learning based approach. In: PACT '10

\bibitem[{Wang and O'boyle(2013)}]{wang2013using}
Wang Z, O'boyle MF (2013) Using machine learning to partition streaming
  programs. ACM TACO

\bibitem[{Wang et~al.(2014{\natexlab{a}})Wang, Tournavitis, Franke, and
  O'Boyle}]{wang2014integrating}
Wang Z, Tournavitis G, Franke B, O'Boyle MFP (2014{\natexlab{a}}) Integrating
  profile-driven parallelism detection and machine-learning-based mapping. ACM
  TACO

\bibitem[{Wang et~al.(2014{\natexlab{b}})}]{DBLP:journals/taco/WangGO14}
Wang Z, et~al. (2014{\natexlab{b}}) Automatic and portable mapping of data
  parallel programs to opencl for gpu-based heterogeneous systems. {ACM TACO}

\bibitem[{Wen et~al.(2014)Wen, Wang, and O'Boyle}]{wen2014smart}
Wen Y, Wang Z, O'Boyle MFP (2014) Smart multi-task scheduling for opencl
  programs on cpu/gpu heterogeneous platforms. In: HiPC '14

\bibitem[{Williams et~al.(2009)Williams, Oliker, Vuduc, Shalf, Yelick, and
  Demmel}]{DBLP:journals/pc/WilliamsOVSYD09}
Williams S, Oliker L, Vuduc RW, Shalf J, Yelick KA, Demmel J (2009)
  Optimization of sparse matrix-vector multiplication on emerging multicore
  platforms. Parallel Computing

\bibitem[{Yuan et~al.(2019)Yuan, Ren, Gao, Tang, and Wang}]{yuan2019using}
Yuan L, Ren J, Gao L, Tang Z, Wang Z (2019) Using machine learning to optimize
  web interactions on heterogeneous mobile systems. IEEE Access
  7:139394--139408

\bibitem[{Zhang(2015)}]{DBLP:conf/hotchips/Zhang15}
Zhang C (2015) Mars: {A} 64-core armv8 processor. In: HotChips

\bibitem[{Zhang et~al.(2018)}]{ipdpsz18}
Zhang P, et~al. (2018) Auto-tuning streamed applications on intel xeon phi. In:
  IPDPS

\end{thebibliography}
